\documentclass[journal,draftcls,onecolumn,12pt]{IEEEtran}

\usepackage{graphicx,cite,epsfig,amssymb,amsmath,algorithm,algorithmic,subfig,subfloat,CJK}
\usepackage{booktabs,longtable,multirow}

\makeatletter
\def\ps@headings{%
\def\@oddhead{\mbox{}\scriptsize\rightmark \hfil \thepage}%
\def\@evenhead{\scriptsize\thepage \hfil \leftmark\mbox{}}%
\def\@oddfoot{}%
\def\@evenfoot{}}

\newcommand{\Rmnum}[1]{\expandafter\@slowromancap\romannumeral #1@}
\makeatother
\begin{document}

\title{Novel $16$-QAM and $64$-QAM Near-Complementary Sequences with Low PMEPR in OFDM Systems}

\author{Tao Jiang, \textit{Senior Member, IEEE}, Chunxing Ni, and Yuance Xu \\
\thanks{This work was supported in part by National Science Foundation (NSFC) for Distinguished Young Scholars of China with Grant number 61325004, the NSFC with Grants 61172052 and 60872008, National High Technology Development 863 Program of China under Grants 2015AA01A710 and 2014AA01A704, the Key Project of Hubei Province in China with Grant 2015BAA074, and the China Scholarship Council (CSC). }
\thanks{Tao Jiang (corresponding author) and Chunxing Ni are with the School of Electronic Information and Communications, Huazhong University of Science and Technology, Wuhan, 430074, China (e-mail: tao.jiang@ieee.org; chunxingni@hust.edu.cn).}
\thanks{Yuance Xu is with Wuhan Posts and Telecommunications Academy of Sciences, Wuhan, 430074, China (e-mail: yuancexu@gmail.com
).

}
}

\date{\today}
\renewcommand{\baselinestretch}{1.2}
\thispagestyle{empty} \maketitle \thispagestyle{empty}
\setcounter{page}{1}

\begin{abstract}
In this paper, we firstly propose a novel construction of $16$-quadrature amplitude modulation (QAM) near-complementary sequences with low peak-to-mean envelope power ratio (PMEPR) in orthogonal frequency division multiplexing (OFDM) systems. The proposed $16$-QAM near-complementary sequences can be constructed by utilizing novel nonlinear offsets, where the length of the sequences is $n=2^m$. The family size of the newly constructed $16$-QAM near-complementary sequences is $8\times (\frac{m!}{2})\times 4^{m+1}$, and the PMEPR of these sequences is proven to satisfy ${\textrm{PMEPR}}\leq 2.4$. Thus, the proposed construction can generate a number of $16$-QAM near-complementary sequences with low PMEPR, resulting in the improvement of the code rate in OFDM systems. Furthermore, we also propose a novel construction of $64$-QAM near-complementary sequences with low PMEPR, which is the first proven construction of $64$-QAM near-complementary sequences. The PMEPRs of two types of the proposed $64$-QAM near-complementary sequences are proven to satisfy that ${\textrm{PMEPR}}\leq 3.62$ or ${\textrm{PMEPR}}\leq 2.48$, respectively. The family size of the newly constructed $64$-QAM near-complementary sequences is $64\times (\frac{m!}{2})\times 4^{m+1}$.
\end{abstract}


\section{Introduction} \label{Introduction}
As an attractive physical layer technology for wireless communications, orthogonal frequency division multiplexing (OFDM) has been applied in many wireless communication standards, due to its considerable
high spectrum efficiency, high power efficiency, multipath delay spread tolerance, and immunity to the frequency selective fading channels~\cite{MS09,Tao08}. However, one major drawback of OFDM is the high peak-to-mean envelope power ratio (PMEPR) of transmitted OFDM signals. Since the high power amplifier (HPA) utilized in OFDM systems has limited linear range, the OFDM signals with high PMEPR will be seriously clipped and nonlinear distortion will be introduced, resulting in serious degradation of the bit error rate (BER) performance~\cite{XXX_000}. Moreover, the high PMEPR leads to the out-of-band radiation, which causes serious adjacent channel interferences~\cite{XXX_001}. Therefore, in order to significantly improve the energy efficiency and reduce communication overhead of OFDM systems, it is necessary to reduce the PMEPR of OFDM systems.

One promising approach for the PMEPR reduction in OFDM systems is to use Golay complementary sequences as codewords~\cite{Davis99}, which can be constructed by co-sets of the classical Reed-Muller codes~\cite{KG00}. The upper-bound of PMEPR for the Golay complementary sequences can be restricted to be less than $2$, which means significant PMEPR reduction for OFDM signals. However, the Golay complementary sequences have extremely low code rate, which makes this coding approach impractical in OFDM systems. For example, the code rate of the $16$-QAM Golay complementary sequences constructed in~\cite{Chong03} is $2.7925$, $1.8962$, $0.4513$ for $n=4$, $8$, $64$, respectively, where the code rate of code $C$ with length $n$ is defined as $R(C)=\frac{\log_{2} |C|}{n}$ bits/symbol~\cite{Lee10}. Moreover, when the length of the sequences increases, the code rate of Golay complementary sequences turns to be extremely low. Thus, the low code rate is the bottleneck of Golay complementary sequences. The most important objective of the coding approach is to improve the code rate of Golay complementary sequences, which means that more and more sequences with low PMEPR should be constructed. Therefore, researchers have proposed another type of sequences: near-complementary sequences, which generalize the Golay complementary sequences, in order to improve the code rate with a slight loss on the PMEPR performance \cite{Schmidt06,Yu11}.

As discussed above, the near-complementary sequences are expected to produce more sequences than Golay complementary sequences at the cost of an increase of the PMEPR bound. Thus, near-complementary sequences can offer low PMEPR with satisfactory code rate, and they are promising to be implemented in OFDM systems for the PMEPR reduction. The main objective of near-complementary sequences is to develop novel constructions of the near-complementary sequences to enlarge their family size. A framework for the construction of $2^h$-phase shift keying (PSK) near-complementary sequences have been proposed in~\cite{Yu11}, which can achieve low PMEPR in OFDM systems. Since $2^h$-quadrature amplitude modulation (QAM) is more widely used in OFDM systems, different constructions of $16$-QAM near-complementary sequences have been proposed in~\cite{BT03, Lee06, Lee10, Li08}, and these $16$-QAM near-complementary sequences can be utilized to significantly reduce the PMEPR in OFDM systems.

In this paper, we propose a novel construction of $16$-QAM near-complementary sequences to enlarge the family size of the $16$-QAM near-complementary sequences. The family size of the newly constructed $16$-QAM near-complementary sequences is $8\times (\frac{m!}{2})\times 4^{m+1}$. Moreover, the proposed $16$-QAM near-complementary sequences satisfy that ${\textrm{PMEPR}}\leq 2.4$, which means a significant PMEPR reduction in OFDM systems.

Furthermore, to our best knowledge, all existing researches have not proposed a construction of $64$-QAM near-complementary sequences, which has been a difficult problem for years. Therefore, we develop a novel construction of $64$-QAM near-complementary sequences with a low PMEPR upper-bound in this paper, and it is the first proven construction for the $64$-QAM near-complementary sequences. Two types of the proposed $64$-QAM near-complementary sequences satisfy that ${\textrm{PMEPR}}^1\leq 3.62$ or ${\textrm{PMEPR}}^2\leq 2.48$, respectively. With the proposed construction, the family size of the newly constructed $64$-QAM near-complementary sequences is $64\times (\frac{m!}{2})\times 4^{m+1}$.

The contributions of this paper are as follows: (1) A novel construction of $16$-QAM near-complementary sequences is proposed, and the family size of sequences with low PMEPR can be enlarged with the help of the proposed construction. (2) A construction of $64$-QAM near-complementary sequences is proposed, which is the first proven construction for the $64$-QAM near-complementary sequences.

The outline of this paper is given as follows. The definitions and notations are presented in Section~\ref{PRELIMINARIES}. The main results on the proposed constructions of 16-QAM and 64-QAM near complementary sequences are given in Section~\ref{MAIN RESULTS}. In Section~\ref{NUMERICAL RESULTS}, we demonstrate the numerical performances of the proposed sequences. Finally, conclusions are drawn in Section~\ref{CONCLUSIONS}.

\section{PRELIMINARIES} \label{PRELIMINARIES}
A summary of basic definitions and key notations of this paper are included in this section.

--- $q$ is an arbitrary even positive integer, $m$, $n$ and $h$ are positive integers.

--- ${Z_q}$ is a ring of integer modulo ${q}$.

--- $Z_q^m$ is an $m-$dimensional vector space where each component is an element in ${{\rm{Z}}_{{q}}}$.

--- $\zeta={e^{ 2j\pi /q}}$ is a primitive $q$th root of unity where $q$ is an even positive integer, $j=\sqrt {- 1}$.

--- $\zeta=e^{ j\pi /2}=j$ when $q=4$.

--- $\mathbb{C}$ denotes the set of complex numbers.

--- $\{ \pi (0),\pi (1), \ldots ,\pi (m - 1)\}$ denotes an arbitrary permutation of the set $\{0,1,\ldots,m-1\}$.

--- $A_i$ denotes the $i$th element of the sequences generated from the function $A(\underline{x})$, where $\underline{x} = ({x_{\pi (0)}},{x_{\pi (1),}}\ldots,{x_{\pi (m - 1)}})$.

\subsection{Peak-to-Mean Envelope Power Ratio (PMEPR)}

In OFDM systems, the transmitted time-domain OFDM signal of the input sequence $\mathbf{A} = ({A_0},{A_1},\ldots,{A_{n - 1}})$ can be obtained by
\begin{align} \label{eq 15a}
{S_\mathbf{A}}(t) = \sum\limits_{i = 1}^{n - 1} {{A_i}{e^{2j\pi {w_i}t}}}, ~0\leq t< T,
\end{align}
where $n$ is the number of subcarriers, ${{w_i}}$ is the frequency of the $i$th carrier, and $T$ is period of the OFDM data block. To ensure orthogonality, the carrier frequencies are related by ${w_i} = {w_0} + i{w_s} {\rm{~}}(i = 0,1,\ldots,n - 1)$, where ${w_0}$ is the smallest carrier frequency and ${w_s}$ is the spacing of the frequencies.

%

\emph{Definition 1 (Instantaneous Envelope Power)}: According to (\ref{eq 15a}), the instantaneous envelope power of the time-domain OFDM signal ${S_\mathbf{A}}(t)$ is defined as
\begin{equation}\label{eq iep}
\begin{array}{l}
{P_\mathbf{A}}(t) = {\left| {{S_\mathbf{A}}(t)} \right|^2} = (\sum\limits_{i = 0}^{n - 1} {{A_i}{e^{2j\pi {w_i}t}}} )(\sum\limits_{k = 0}^{n - 1} {A_k^ * {e^{ - 2j\pi {w_k}t}}} ) \\
{\rm{~~~~~~~}}= \sum\limits_{i = 0}^{n - 1} {\sum\limits_{k = 0}^{n - 1} {{A_i}A_k^ * } } {e^{2j\pi (i - k){w_s}t}}.
\end{array}
\end{equation}

\emph{Definition 2 (PEP of a codeword)}: The peak envelope power (PEP) of a codeword $\mathbf{A}$ is defined as
\begin{equation}\label{eq pep}
{\textrm{PEP}}(\mathbf{A})= \mathop {\sup }\limits_{t \in [0,T]} {{P_\mathbf{A}}(t)},
\end{equation}
where $\sup (\cdot)$ denotes the supremum.

\emph{Definition 3 (PMEPR of a code)}: The PMEPR of a code $\mathcal{C}$ is defined as the ratio of the PEP to the average mean power of code $\mathcal{C}$ over all OFDM signals generated from a codebook $\mathcal{C}$~\cite{Lee10}, i.e.,
\begin{equation}\label{eq pmepr}
{\textrm{PMEPR}}(\mathcal{C})= \mathop {\max }\limits_{\mathbf{A} \in \mathcal{C}} {\textrm{PEP}}(\mathbf{A})/P_{av}(\mathcal{C}),
\end{equation}
where $P_{av}(\mathcal{C})$ denotes the average mean power of the code $\mathcal{C}$. Moreover, $P_{av}(\mathcal{C})$ is a constant for codebook $\mathcal{C}$.

\subsection{Generalized Boolean Functions}

\emph{Definition 4 (Generalized Boolean Function)}: A generalized Boolean function $f$ is defined by a mapping $f: Z_2^m  \to Z_q$, and it can be written in an algebraic normal form \cite{Schmidt06} as
\begin{equation}\label{eq 11a}
f(x) = f(x_0 ,x_1 ,\ldots,x_{m - 1} ) = \sum\limits_{p = 0}^{2^m  - 1} {c_p \prod\limits_{l = 0}^{m - 1} {x_l^{i_l } } },
\end{equation}
where sequence $x=(x_0,x_1,\ldots,x_{m-1})$, ${c_p}\in{Z_q}$, and ${i_l}\in{Z_2}~(l=0,1,\cdots,m-1)$ can be obtained from a binary representation of $p = \sum\limits_{l = 0}^{m - 1} {i_l 2^l }$.

A generalized Boolean function can also be represented by sequences of length $2^m$. The $Z_q$-valued sequence associated with $f$ is defined as
\begin{equation}\label{eq 12a}
\psi (f) = (f_0,f_1,\ldots,f_{2^m  - 1} ),
\end{equation}
where $f_i=f(i_0,i_1,\ldots,i_{m-1})$, and $(i_0,i_1,\ldots,i_{m-1})$ is the binary representation of $0 \le i \le 2^m-1$. Moreover, the sequence $\Psi(f)$ is defined as the polyphase sequence associated with $f$ as
\begin{equation} \label{eq 13a}
\Psi(f)=(\zeta ^{f_0 },\zeta ^{f_1 },\ldots,\zeta ^{f_{2^m  - 1} } ),
\end{equation}

Let $f:Z_2^k \to {Z_q}$ be a generalized Boolean function. The sequence $\phi (f)$ of length $({4^k} + 2)/3$ is defined as
\begin{align} \label{eq 14a}
\phi (f) = \left\{ \begin{array}{l}
 {\zeta ^{f({x_0},{x_1},\ldots,{x_{k - 1}})}}, \textrm{at position}{\rm{~}}\sum\limits_{a = 0}^{k - 1} {{x_a}{2^{2a}}},  \\
 0, {\rm{~~~~~~~~~~~~~~~}}\textrm{otherwise},  \\
 \end{array} \right.
\end{align}
where ${x_0},{x_1},\ldots,{x_{k-1}}$ range over $Z_2^k$.

\subsection{Golay Complementary Sequences}
\emph{Definition 5 (Polyphase Sequence)}: For a complex sequence $\mathbf{A} = ({A_0},{A_1},\ldots,{A_{n - 1}})$, if ${A_i} = {\zeta ^{{a_i}}}$, and ${a_i} \in {Z_{{q}}}$ for $i=0,1,\ldots,n-1$, then $\mathbf{A}$ is called a polyphase sequence.

\emph{Definition 6 (Aperiodic Auto-Correlation Function)}: The aperiodic auto-correlation function of sequence $\mathbf{A} = ({A_0},{A_1},\ldots,{A_{n - 1}})$ is defined as
\begin{align} \label{eq 16a}
{C_\mathbf{A}}(u) = \left\{ \begin{array}{l}
 \sum\limits_{i = 0}^{n - 1 - u} {{A_i}A_{i + u}^*,{\rm{~}}0 \le u < n},  \\
 \sum\limits_{i = 0}^{n - 1 + u} {{A_{i - u}}A_i^*,{\rm{~}}- n < u < 0},  \\
 0,{\rm{~~~~~~~~~~~~~~~~~}} \textrm{otherwise}.  \\
\end{array} \right.
\end{align}
Note that, ${C_\mathbf{A}}(u)$ is conjugate-symmetric, i.e., ${C_\mathbf{A}}( - u) = {{C^*}_\mathbf{A}}(u)$, and ${{C^*}_\mathbf{A}}(u)$ is the complex conjugate of ${C_\mathbf{A}}(u)$.

\emph{Definition 7 (Operator $\star$)}: For two sequences $\mathbf{A},{\rm{~}}\mathbf{B }\in \mathbb{C}^n$, the operator $\star$ is defined as
\begin{align} \label{eq 17a}
\mathbf{A} \star \mathbf{B} = \sum\limits_{u = 1 - n}^{n - 1} {\left| {{C_\mathbf{A}}(u) + {C_\mathbf{B}}(u)} \right|}.
\end{align}

If $\mathbf{A}$ and $\mathbf{B}$ are polyphase sequences of length $n$, we have
\begin{equation} \label{eq 18a}
\begin{split}
\mathbf{A} \star \mathbf{B} &= 2\sum\limits_{u = 0}^{n - 1} {\left| {{C_\mathbf{A}}(u) + {C_\mathbf{B}}(u)} \right|} \\
&= 2n + 2\sum\limits_{u = 1}^{n - 1} {\left| {{C_\mathbf{A}}(u) + {C_\mathbf{B}}(u)} \right|}.
\end{split}
\end{equation}

\emph{Definition 8 (Golay Complementary Sequence)}: If $\mathbf{A} \star \mathbf{B} = 2n$ (i.e., ${C_\mathbf{A}}(u) + {C_\mathbf{B}}(u) = 0$ for all $u \neq 0$), the pair of polyphase sequences $(\mathbf{A},\mathbf{B})$ is called a Golay complementary pair, and the sequence $\mathbf{A}$ or $\mathbf{B}$ is called a Golay complementary sequence.

According to \eqref{eq iep}, \eqref{eq pmepr} and \emph{Theorem 2} in~\cite{Schmidt06}, the PMEPR of $\mathbf{A}$ and $\mathbf{B}$ is at most $\mathbf{A} \star \mathbf{B}$, i.e.,
\begin{equation}\label{eq pmepr2}
{\textrm{PMEPR}}(\mathbf{A})\leq (\mathbf{A} \star \mathbf{B})/n.
\end{equation}

When $\mathbf{A}$ is a polyphase Golay complementary sequence, then ${C_\mathbf{A}}(u) + {C_\mathbf{B}}(u) = 0$ for all $u \neq 0$. Substituting (\ref{eq 18a}) into (\ref{eq pmepr2}), we can obtain
\begin{equation}\label{eq pmepr3}
{\textrm{PMEPR}}(\mathbf{A})\leq (\mathbf{A} \star \mathbf{B})/n=2.
\end{equation}
Therefore, the PMEPR of Golay complementary sequences is at most $2$, which provides significant PMEPR reduction.



\subsection{Near-Complementary Sequences}

\emph{Definition 9 (Near-Complementary Sequence)}: We call a pair of polyphase sequences $(\mathbf{A},\mathbf{B})$ as a near-complementary pair if $2n \le \mathbf{A} \star \mathbf{B} \ll 2{n^2}$, and sequence $\mathbf{A}$ or $\mathbf{B}$ is called as a near-complementary sequence \cite{Schmidt06}.

\subsection{Decompositions of QAM Symbols}

To our best knowledge, any $16$-QAM symbol can be decomposed into a pair of quadrature phase shift keying (QPSK) symbols~\cite{Chong03}, because any point on the $16$-QAM constellation can be written as
\begin{equation}\label{eq 16qam}
s(u,v) = \gamma ({r_1}{\zeta ^u} + {r_2}{\zeta ^v}),
\end{equation}
where $\gamma = {e^{j\pi /4}}$, $\zeta = e^{ j\pi /2} = j$, $u,v \in Z_4$, ${r_1} = 2/\sqrt 5$ and ${r_2} = 1/\sqrt 5$ are required for the constellation to have unit average energy.

Similarly, a $64$-QAM symbol could be constructed with three QPSK symbols as~\cite{Lee06}
\begin{equation}\label{eq 64qam}
q(u,v,w) = \gamma ({a_1}{\zeta ^u} + {a_2}{\zeta ^v} + {a_3}{\zeta ^w}),
\end{equation}
where $u,v,w \in {Z_4}$. To maintain its average squared magnitude, ${a_1} = \frac{4}{{\sqrt {21} }}, {a_2} = \frac{2}{{\sqrt {21} }}, {a_3} = \frac{1}{{\sqrt {21} }}$.

\section{Main Results} \label{MAIN RESULTS}

\subsection{Proposed Construction of $16$-QAM Near-Complementary Sequences}\label{MAIN RESULTS_1}

In this subsection, we propose a novel construction of $16$-QAM near-complementary sequences in \emph{Theorem 1}.

\emph{Theorem 1}: Suppose $m > 2$ and let $D,E,D^{'},E^{'}: Z_2^m \to {Z_4}$, be four generalized Boolean functions of variable $\underline{x} = ({x_{\pi (0)}},{x_{\pi (1),}}\ldots,{x_{\pi (m - 1)}}) \in Z_2^m$, where $\{ \pi (0),\pi (1), \ldots ,\pi (m - 1)\}$ denotes an arbitrary permutation of the set $\{0,1,\ldots,m-1\}$. Let
\begin{equation}\label{eq_qam16_1}
\begin{split}
D(\underline{x}) = &2\sum\limits_{l = 0}^{m - 2} {{x_{\pi (l)}}{x_{\pi (l + 1)}} + \sum\limits_{l = 0}^{m - 1} {{c_l}{x_{\pi (l)}} + c}}.\\
E(\underline{x}) = &D(\underline{x}) + 2{x_{\pi (0)}}{x_{\pi (1)}} + {d_1}{x_{\pi (0)}} + {d_2}{x_{\pi (1)}} + {d_3},\\
{D^{'}}(\underline{x}) = &D(\underline{x}) + 2{x_{\pi (m - 1)}},\\
{E^{'}}(\underline{x}) = &E(\underline{x}) + 2{x_{\pi (m - 1)}},
\end{split}
\end{equation}
where $c_l, c \in Z_4$, and ${d_1} + 2{d_3} = 2$, $2{d_2} = 2$, ${d_1},{d_2},{d_3} \in {Z_4}$. Define the offset $s({\underline{x}})$ as
\begin{equation}\label{eq s}
s({\underline{x}}) = E(\underline{x}) - D(\underline{x}) = 2{x_{\pi (0)}}{x_{\pi (1)}} + {d_1}{x_{\pi (0)}} + {d_2}{x_{\pi (1)}} + {d_3}.
\end{equation}

Then, the proposed $16$-QAM sequence
\begin{equation}\label{eq_code}
{\textrm{H}(\underline{x})} = \gamma ({r_1}{\zeta ^{{D(\underline{x})}}} + {r_2}{\zeta ^{{E(\underline{x})}}})
\end{equation}
is a near-complementary sequence and ${\rm{PMEPR}}({\textrm{H}}) \le 2.4$, where $\gamma = {e^{j\pi /4}}$, $\zeta = e^{ j\pi /2} = j$, ${r_1} = 2/\sqrt 5$ and ${r_2} = 1/\sqrt 5$.

\emph{Proof}: See APPENDIX \uppercase\expandafter{\romannumeral1}.

\emph{Corollary 1}: The family size of the construction in \emph{Theorem 1} is $8 \times (\frac{{m!}}{2}) \times {4^{m + 1}}$, and the length of the sequences is $n={2^m}$.

\emph{Proof}: It can be seen in \emph{Theorem 1} that the number of the offsets is $8$. According to \cite{Davis99} and \emph{Theorem 1}, we can calculate the family size of this construction to be $8 \times (\frac{{m!}}{2}) \times {4^{m + 1}}$, and the length of the sequences $n={2^m}$.

Then, we give an example of \emph{Theorem 1} as follows.

\emph{Example 1}: Let $m=3$, ${d_1}=0$, ${d_2}=1$ and ${d_3}=1$, $D(\underline{x})=\{0,1,1,0,1,2,0,3\}$, $E(\underline{x})=\{1,2,3,2,2,3,0,3\}$, and ${\textrm{H}(\underline{x})}=\{\frac{1}{{\sqrt {10} }}(-1+3j),\frac{1}{{\sqrt {10} }}(-3-j),\frac{1}{{\sqrt {10} }}(1-j),\frac{1}{{\sqrt {10} }}(-1-j),\frac{1}{{\sqrt {10} }}(-3-j),\frac{1}{{\sqrt {10} }}(1-3j),\frac{1}{{\sqrt {10} }}(3+3j),\frac{1}{{\sqrt {10} }}(3-3j)\}$. Thus, we can obtain that ${\rm{PMEPR}}({\textrm{H}}) \approx 2.1 \leq 2.4$.

\subsection{Proposed Constructions of $64$-QAM Near-Complementary Sequences}
Based on the discussion in Subsection \ref{MAIN RESULTS_1}, we propose a novel construction of $64$-QAM near-complementary sequences in this subsection, which is presented in~\emph{Theorem 2}.

\emph{Theorem 2}: Suppose $m>2$, and denote $D,F,G,D^{'},F^{'},G^{'}: Z_2^m \to {Z_4}$ as six generalized Boolean functions of variable $\underline{x} = ({x_{\pi (0)}},{x_{\pi (1),}}\ldots,{x_{\pi (m - 1)}}) \in Z_2^m$, where $\{ \pi (0),\pi (1), \ldots ,\pi (m - 1)\}$ denotes an arbitrary permutation of the set $\{0,1,\ldots,m-1\}$. Let
\begin{equation}\label{eq s3}
\begin{split}
D(\underline{x}) &= 2\sum\limits_{l = 0}^{m - 2} {{x_{\pi (l)}}{x_{\pi (l + 1)}} + \sum\limits_{l = 0}^{m - 1} {{c_l}{x_{\pi (l)}} + c}}.\\
F(\underline{x}) &= D(\underline{x}) + s^{(1)}(\underline{x}),\\
G(\underline{x}) &= D(\underline{x}) + s^{(2)}(\underline{x}),\\
{D^{'}}(\underline{x}) &= D(\underline{x}) + 2{x_{\pi (m - 1)}}, \\
{F^{'}}(\underline{x}) &= F(\underline{x}) + 2{x_{\pi (m - 1)}}, \\
{G^{'}}(\underline{x}) &= G(\underline{x}) + 2{x_{\pi (m - 1)}}, \\
\end{split}
\end{equation}
where $c_k \in Z_4$, $c \in Z_4$. Then, the $64$-QAM sequence
\begin{equation}\label{eq_qam64}
{\textrm{J}(\underline{x})} = \gamma ({a_1}{\zeta ^{{D(\underline{x})}}} + {a_2}{\zeta ^{{F(\underline{x})}}} + {a_3}{\zeta ^{{G (\underline{x})}}})
\end{equation}
is a near-complementary sequence, where $\gamma = {e^{j\pi /4}}$, $\zeta = e^{ j\pi /2} = j$, ${a_1} = \frac{4}{{\sqrt {21} }}, {a_2} = \frac{2}{{\sqrt {21} }}$, ${a_3} = \frac{1}{{\sqrt {21} }}$ for each one of the following offset pairs.

\emph{Type 1}:
\begin{equation}
\begin{split}
&s^{(1)}(\underline{x}) =  {h_1}{x_{\pi (0)}}  + {h_3},  \\
&s^{(2)}(\underline{x}) = 2{x_{\pi (0)}}{x_{\pi (1)}} + {d_1}{x_{\pi (0)}} + {d_2}{x_{\pi (1)}} + {d_3}, \\
&{\textmd{for}}~{h_1} + 2{h_3} = 0, {h_1},{h_3} \in {Z_4}.
\end{split}
\end{equation}

\emph{Type 2}:
\begin{equation}
\begin{split}
&s^1(\underline{x}) = 2{x_{\pi (0)}}{x_{\pi (1)}} + {d_1}{x_{\pi (0)}} + {d_2}{x_{\pi (1)}} + {d_3}, \\
&s^2(\underline{x}) = 2{x_{\pi (0)}}{x_{\pi (1)}} + {h_1}{x_{\pi (0)}} + {h_2}{x_{\pi (1)}} + {h_3},\\
&{\textmd{for}}~{h_2}={d_2} + 2, {h_1} + 2{h_3} = 2, {h_1},{h_2},{h_3} \in {Z_4}.
\end{split}
\end{equation}

Denote PMEPR$^1$ and PMEPR$^2$ as the PMEPR of 64-QAM sequences of \emph{Type 1} and \emph{Type 2}, respectively, then, $\textrm{PMEPR}^1 \leq 3.62$ and $\textrm{PMEPR}^2 \leq 2.48$.

\emph{Proof of Theorem 2}: See APPENDIX \uppercase\expandafter{\romannumeral2}.

\emph{Corollary 2}: The family size of the $64$-QAM near-complementary sequences in \emph{Theorem 2} is $64 \times \frac{{m!}}{2} \times 4^{m+1}$, where the length of the sequences is $2^m$.

\emph{Proof}: It is obvious that 64 offsets can be constructed from \emph{Type 1} and \emph{Type 2}. According to \cite{Davis99}, the family size of the $64$-QAM near-complementary sequences proposed in \emph{Theorem 2} is $64 \times \frac{{m!}}{2} \times 4^{m+1}$.

An example of \emph{Theorem 2} is given as follows.

\emph{Example 2}: Let $m=3$, ${d_1}=0$, ${d_2}=1$, ${d_3}=1$, ${h_1}=0$ and ${h_3}=0$, which are included in \emph{Type 2}. Therefore, $D(\underline{x})=\{0,1,1,0,1,2,0,3\}$, $E(\underline{x})=\{0,1,1,0,1,2,0,3\}$, $F(\underline{x})=\{1,2,3,2,2,3,0,3\}$, and ${\textrm{H}(\underline{x})}=\{\frac{1}{{\sqrt {42} }}(5 + 7j),\frac{1}{{\sqrt {42} }}( - 7 + 5j),\frac{1}{{\sqrt {42} }}( - 5 + 5j),\frac{1}{{\sqrt {42} }}(5 + 5j),\frac{1}{{\sqrt {42} }}( - 7 + 5j),\frac{1}{{\sqrt {42} }}( - 5 - 7j),\frac{1}{{\sqrt {42} }}(7{\rm{ + }}7j),\frac{1}{{\sqrt {42} }}(7{\rm{ - }}7j)\}$. Thus, we can obtain that ${\rm{PMEPR}}({\textrm{H}}) \approx 3.5 \leq 3.62$.

\section{NUMERICAL RESULTS} \label{NUMERICAL RESULTS}
\subsection{PMEPR Reduction}
As discussed in Section \ref{MAIN RESULTS}, the PMEPR of the proposed $16$-QAM near-complementary sequences has been proven to satisfy that ${\textrm{PMEPR}}\leq 2.4$. Moreover, the PMEPRs of the two types of the proposed $64$-QAM near-complementary sequences satisfy that ${\textrm{PMEPR}}^1\leq 3.62$ and ${\textrm{PMEPR}}^2\leq 2.48$, respectively. In this subsection, some numerical results have been conducted to evaluate the PMEPR reduction performances of the proposed $16$-QAM near-complementary sequences and the proposed $64$-QAM near-complementary sequences. The complementary cumulative distribution function (CCDF) is employed in the simulations to measure the PMEPR reduction performances of the proposed $16$-QAM and $64$-QAM near-complementary sequences. The CCDF is defined as the probability that the PMEPR exceeds a given threshold $\textmd{PMEPR}_0$, i.e.,
\begin{equation}\label{eq_add103}
\textmd{CCDF}=Pr\left\{\textmd{PMEPR}>\textmd{PMEPR}_0\right\}.
\end{equation}

\begin{figure}[!h]
 \begin{center}
 \includegraphics[width=3.5in]{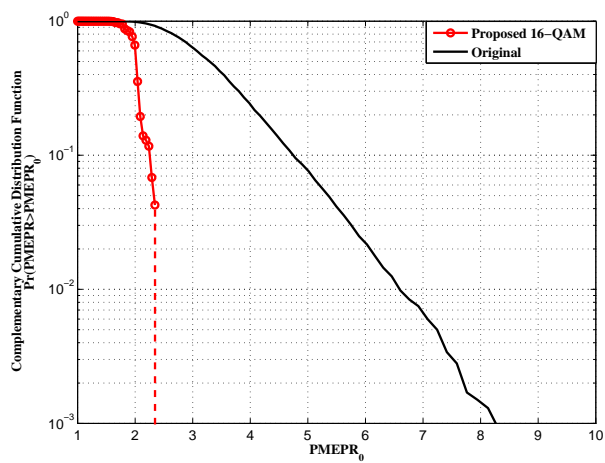}
 \end{center}
 \caption{PMEPR reduction of the proposed $16$-QAM near-complementary sequences with $m=4$ and $n=16$.}
 \label{f1}
\end{figure}

Fig.~\ref{f1} depicts the the PMEPR reduction performance of the proposed $16$-QAM near-complementary sequences with $m=4$ and $n=16$. The ``Original'' curve shows the PMEPR reduction performance of the conventional OFDM signals without PMEPR reduction. Seen from Fig.~\ref{f1}, the PMEPR of the proposed $16$-QAM near-complementary sequences is less than or equal to $2.4$, which is consistent with our proof in Section \ref{MAIN RESULTS}. Therefore, the proposed $16$-QAM near-complementary sequences can significantly reduce the PMEPR in OFDM systems.

\begin{figure}[!h]
 \begin{center}
 \includegraphics[width=3.5in]{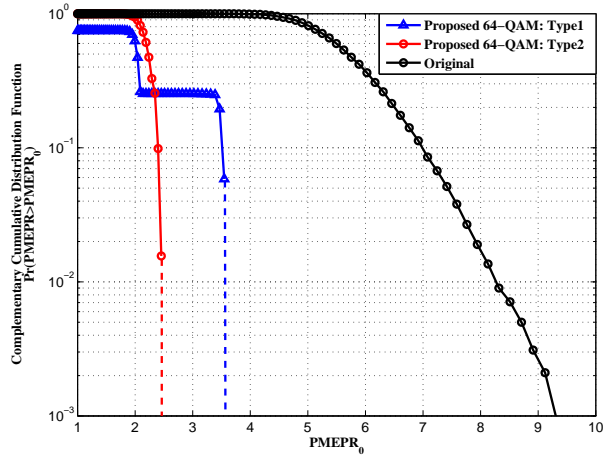}
 \end{center}
 \caption{PMEPR reduction of the proposed $64$-QAM near-complementary sequences with $m=4$ and $n=16$.}
 \label{f2}
\end{figure}

Fig.~\ref{f2} shows the the PMEPR reduction performance of the proposed $64$-QAM near-complementary sequences with $m=4$ and $n=16$. The ``Original'' curve shows the PMEPR reduction performance of the conventional OFDM signals without PMEPR reduction. Seen from Fig.~\ref{f2}, the proposed $64$-QAM near-complementary sequences in \emph{Type 1} satisfy that ${\textrm{PMEPR}}^1\leq 3.62$, and the proposed $64$-QAM near-complementary sequences in \emph{Type 2} satisfy that ${\textrm{PMEPR}}^2\leq 2.48$. The simulation results are consistent with the discussion in Section \ref{MAIN RESULTS}. Therefore, the proposed $64$-QAM near-complementary sequences can significantly reduce the PMEPR in OFDM systems.

In summary, the proposed $16$-QAM and $64$-QAM near-complementary sequences both offer significant PMEPR reductions in OFDM systems.

\subsection{Family Size}
In this subsection, some numerical results have been presented in TABLE \ref{n1} and TABLE \ref{n2} to show family sizes of the proposed $16$-QAM and $64$-QAM near-complementary sequences. For comparison, we also present the family size of the $16$-QAM near-complementary sequences proposed in \cite{Lee10}. Furthermore, since the proposed construction of $64$-QAM near-complementary sequences is the first proven construction, we cannot compare it with any existing $64$-QAM near-complementary sequences. In the numerical results, the length of the sequences is $n=2^m$, where $m$ is a positive integer and $m>2$. Moreover, we denote ${\mathcal{N}^1_{16}}$, ${\mathcal{N}^2_{16}}$, and ${\mathcal{N}^1_{64}}$ as the numbers of the proposed $16$-QAM near-complementary sequences, the $16$-QAM near-complementary sequences in \cite{Lee10}, and the proposed $64$-QAM near-complementary sequences, respectively.

\begin{table}[!h]
\renewcommand{\arraystretch}{1.3}
\caption{NUMBERS OF $16$-QAM SEQUENCES} \label{n1} \centering
\begin{tabular}{|c|c|c|c|}
\hline
\multirow{2}{*}{Numbers}   &  \multicolumn{3}{|c|}{The length of sequences}     \\
\cline{2-4}
& $m=3,n=8$ & $m=4,n=16$ & $n=2^m$    \\
\hline
${\mathcal{N}^1_{16}}$  &  $6144$ & $98304$ & $8 \times (\frac{{m!}}{2}) \times {4^{m + 1}}$       \\
\hline
${\mathcal{N}^2_{16}}$  & $768$ & $12288$ & $(\frac{{m!}}{2}) \times {4^{m + 1}}$  \\
\hline
\end{tabular}
\end{table}

As shown in TABLE \ref{n1}, it is obvious that the number of the proposed $16$-QAM near-complementary sequences is larger than the $16$-QAM near-complementary in \cite{Lee10}. For example, when $m=4$ and $n=16$, the numbers of the proposed $16$-QAM near-complementary sequences and the $16$-QAM near-complementary sequences in \cite{Lee10} are $98304$ and $12288$, respectively. Thus, the proposed construction of the $16$-QAM near-complementary sequences can enlarge the family size of near-complementary sequences, resulting in more near-complementary sequences with low PMEPR. Therefore, more near-complementary sequences with low PMEPR can be utilized in OFDM systems, resulting in the improvement of the code rate in OFDM systems.

\begin{table}[!h]
\renewcommand{\arraystretch}{1.3}
\caption{NUMBERS OF $64$-QAM SEQUENCES} \label{n2} \centering
\begin{tabular}{|c|c|c|c|}
\hline
\multirow{2}{*}{Numbers}   &  \multicolumn{3}{|c|}{The length of sequences}     \\
\cline{2-4}
& $m=3,n=8$ & $m=123,n=2^{123}$ & $n=2^m$    \\
\hline
${\mathcal{N}^1_{64}}$ & $49152$ & $64 \times \frac{{123!}}{2} \times 4^{124}$ & $64 \times \frac{{m!}}{2} \times 4^{m+1}$   \\
\hline
\end{tabular}
\end{table}

We present the number of the proposed $64$-QAM near-complementary sequences in TABLE \ref{n2}. As shown in TABLE \ref{n2}, the family size of the proposed $64$-QAM near-complementary sequences is $64(\frac{m!}{2})4^{m+1}$. For example, the number of the proposed $64$-QAM near-complementary sequences is $49152$ when $m=3$ and $n=8$. Therefore, the proposed construction of $64$-QAM near-complementary sequences is of great value, and it can be employed to control the PMEPR in OFDM systems.

\section{CONCLUSIONS} \label{CONCLUSIONS}

In this paper, a novel construction of $16$-QAM near-complementary sequences was proposed to reduce the PMEPR in OFDM systems. The family size of the newly constructed $16$-QAM near-complementary sequences is $8\times (\frac{m!}{2})\times 4^{m+1}$, and the PMEPR of the sequences is bounded by $2.4$. Moreover, a construction of $64$-QAM near-complementary sequences was also proposed in this paper, which is the first proven construction of $64$-QAM near-complementary sequences. The family size of the newly constructed $64$-QAM near-complementary sequences is $64\times (\frac{m!}{2})\times 4^{m+1}$, and the PMEPR of the sequences is bounded by $2.48$ or $3.62$. Therefore, the proposed $16$-QAM and $64$-QAM near-complementary sequences offer significant PMEPR reduction.

\begin{center}
APPENDIX \uppercase\expandafter{\romannumeral1}
\end{center}
\begin{center}
The Proof Of \emph{Theorem 1}
\end{center}

The proof of \emph{Theorem 1} consists of two steps: (1) To prove that the proposed $16$-QAM sequence ${\textrm{H}(\underline{x})} = \gamma ({r_1}{\zeta ^{{D(\underline{x})}}} + {r_2}{\zeta ^{{E(\underline{x})}}})$ is a near-complementary sequence; (2) To prove that the PMEPR upper bound of {\textrm{H}(\underline{x})} satisfies that ${\rm{PMEPR}}({\textrm{H}}) \le 2.4$.

Then let us start the proof of \emph{Theorem 1} as follows.

Firstly, let $\underline{i} = ({i_0},{i_1},...,{i_{m - 1}})$ be the binary representation of $i$, i.e., $i = \sum\limits_{k = 0}^{m - 1} {{i_k}{2^{m - k}}}$. Then, recall \eqref{eq_qam16_1} and \eqref{eq s}, we have
\begin{equation}\label{eq_qam16_ap1}
\begin{split}
D(\underline{x}) = &2\sum\limits_{l = 0}^{m - 2} {{x_{\pi (l)}}{x_{\pi (l + 1)}} + \sum\limits_{l = 0}^{m - 1} {{c_l}{x_{\pi (l)}} + c}}.\\
E(\underline{x}) = &D(\underline{x}) + 2{x_{\pi (0)}}{x_{\pi (1)}} + {d_1}{x_{\pi (0)}} + {d_2}{x_{\pi (1)}} + {d_3},\\
{D^{'}}(\underline{x}) = &D(\underline{x}) + 2{x_{\pi (m - 1)}},\\
{E^{'}}(\underline{x}) = &E(\underline{x}) + 2{x_{\pi (m - 1)}},
\end{split}
\end{equation}
and the offset $s({\underline{x}})$ is expressed as
\begin{equation}\label{eq s_ap1}
s({\underline{x}}) = E(\underline{x}) - D(\underline{x}) = 2{x_{\pi (0)}}{x_{\pi (1)}} + {d_1}{x_{\pi (0)}} + {d_2}{x_{\pi (1)}} + {d_3}.
\end{equation}

Then, we define ${\textrm{H}^{'}}(\underline{x})$ as follows:
\begin{equation}\label{eq_ap1}
{\textrm{H}^{'}(\underline{x})} = \gamma ({r_1}{\zeta ^{{{D^{'}}(\underline{x})}}} + {r_2}{\zeta ^{{{E^{'}}(\underline{x})}}}).
\end{equation}

Substituting \eqref{eq_code} into \eqref{eq 16a}, the aperiodic auto-correlation function of the proposed $16$-QAM sequence $\textrm{H}(\underline{x})$ is
\begin{equation}\label{eq aH}
\begin{array}{l}
{C_{\rm{H}}}(u) = \sum\limits_{i = 0}^{n - u - 1} {({r_1}{\zeta ^{{D_i}}} + {r_2}{\zeta ^{{E_i}}})} {({r_1}{\zeta ^{{D_{i+u}}}} + {r_2}{\zeta ^{{E_{i+u}}}})^*} \\
{\rm{~~~~~~~}} = \sum\limits_{i = 0}^{n - u - 1} {[r_1^2{\zeta ^{{D_i} - {D_{i + u}}}} + r_2^2{\zeta ^{{E_i} - {E_{i + u}}}}} {+ {r_1}{r_2}{\zeta ^{{D_i} - {E_{i + u}}}}} \\
{\rm{~~~~~~~~~~~~~~~~~~}} + {r_1}{r_2}{\zeta ^{{E_i} - {D_{i + u}}}}].
\end{array}
\end{equation}

Similarly, substituting \eqref{eq_ap1} into \eqref{eq 16a}, the aperiodic auto-correlation function of $\textrm{H}^{'}(\underline{x})$ is
\begin{equation}\label{eq aHc}
\begin{array}{l}
{C_{{\rm{H^{'}}}}}(u) = \sum\limits_{i = 0}^{n - u - 1} {({r_1}{\zeta ^{{D_i^{'}}}} + {r_2}{\zeta ^{{E_i^{'}}}})} {({r_1}{\zeta ^{{D_{i+u}^{'}}}} + {r_2}{\zeta ^{{E_{i+u}}^{'}}})^*}\\
{\rm{~~~~~~~~}} = \sum\limits_{i = 0}^{n - u - 1} {[r_1^2{\zeta ^{{D_i^{'}} - {D_{i + u}^{'}}}} + r_2^2{\zeta ^{{E_i^{'}} - {E_{i + u}^{'}}}} }{+ {r_1}{r_2}{\zeta ^{{D_i^{'}} - {E_{i + u}^{'}}}}} \\
 {\rm{~~~~~~~~~~~~~~~~~~}}+ {r_1}{r_2}{\zeta ^{{E_i^{'}} - {D_{i + u}^{'}}}}].
\end{array}
\end{equation}

According to \eqref{eq aH} and \eqref{eq aHc}, we have
\begin{equation}\label{eq add16}
\begin{array}{l}
{\rm{~~}}{C_{\rm{H}}}(u) + {C_{{{\rm{H}}^{'}}}}(u) \\
= \sum\limits_{i = 0}^{n - u - 1} {\{ r_1^2[{\zeta ^{{D_i} - {D_{i + u}}}} + {\zeta ^{D_i^{'} - D_{i + u}^{'}}}] }{+ r_2^2[{\zeta ^{{E_i} - {E_{i + u}}}} + {\zeta ^{E_i^{'} - E_{i + u}^{'}}}]  }    \\
{\rm{~~~~~~~}}+{r_1}{r_2}{\zeta ^{{D_i} - {E_{i + u}}}} + {r_1}{r_2}{\zeta ^{{E_i} - {D_{i + u}}}}\\
 {\rm{~~~~~~~}}+ {r_1}{r_2}{\zeta ^{D_i^{'} - E_{i + u}^{'}}} + {r_1}{r_2}{\zeta ^{E_i^{'} - D_{i + u}^{'}}}\}  .
\end{array}
\end{equation}

With \eqref{eq_qam16_ap1}$\thicksim$\eqref{eq s_ap1}, we have
\begin{equation}
{r_1}{r_2}{\zeta ^{{D_i} - {E_{i + u}}}} + {r_1}{r_2}{\zeta ^{{E_i} - {D_{i + u}}}} = {r_1}{r_2}{\zeta ^{{D_i} - {D_{i + u}}}}({\zeta ^{{s_i}}} + {\zeta ^{ - {s_{i + u}}}}), \nonumber
\end{equation}
\begin{equation}
\begin{array}{l}
{\rm{~~~}}{r_1}{r_2}{\zeta ^{D_i^{'} - E_{i + u}^{'}}} + {r_1}{r_2}{\zeta ^{E_i^{'} - D_{i + u}^{'}}} \\
= {r_1}{r_2}{\zeta ^{{D_i} - {D_{i + u}}}} \times {\zeta ^{2{i_{\pi (m - 1)}} - 2{{(i + u)}_{\pi (m - 1)}}}} \times ({\zeta ^{{s_i}}} + {\zeta ^{ - {s_{i + u}}}}), \nonumber
\end{array}
\end{equation}
\begin{equation}
\begin{array}{l}
{\rm{~~~}}{r_1}{r_2}{\zeta ^{({D_i} - {E_{i + u}})}} + {r_1}{r_2}{\zeta ^{{E_i} - {D_{i + u}}}} \\
{\rm{~~~}}+ {r_1}{r_2}{\zeta ^{D_i^{'} - E_{i + u}^{'}}}+ {r_1}{r_2}{\zeta ^{E_i^{'} - D_{i + u}^{'}}} \\
= {r_1}{r_2}{\zeta ^{{D_i} - {D_{i + u}}}}({\zeta ^{{s_i}}} + {\zeta ^{ - {s_{i + u}}}})\\
{\rm{~~~}} \times [1 + {( - 1)^{{i_{\pi (m - 1)}} - {{(i + u)}_{\pi (m - 1)}}}}], \nonumber
\end{array}
\end{equation}
where ${{(i + u)_{\pi (0)}}, {(i + u)_{\pi (1)}},..., {(i + u)_{\pi (m - 1)}}}$ is the binary representation of $i + u$.

Therefore, \eqref{eq add16} can be rewritten as,
\begin{equation}\label{eq Add16}
\begin{array}{l}
{\rm{~~~}}{C_{\rm{H}}}(u) + {C_{{{\rm{H}}^{'}}}}(u) \\
= \sum\limits_{i = 0}^{n - u - 1} {\{ r_1^2[{\zeta ^{{D_i} - {D_{i + u}}}} + {\zeta ^{D_i^{'} - D_{i + u}^{'}}}]}\\
{\rm{~~~}} {+ r_2^2[{\zeta ^{{E_i} - {E_{i + u}}}} + {\zeta ^{E_i^{'} - E_{i + u}^{'}}}]  } \\
{\rm{~~~}}+ {r_1}{r_2}{\zeta ^{{D_i} - {D_{i + u}}}}({\zeta ^{{s_i}}} + {\zeta ^{ - {s_{i + u}}}}) \\
{\rm{~~~}}\times [1 + {( - 1)^{{i_{\pi (m - 1)}} - {{(i + u)}_{\pi (m - 1)}}}}]\}.
\end{array}
\end{equation}

To prove \emph{Theorem 1}, the following lemma is needed.

\emph{Lemma 1}: We can obtain that
\begin{equation}\label{eq lemma1}
\begin{array}{l}
\sum\limits_{u = 1-n}^{n - 1} {\sum\limits_{i = 0}^{n - u - 1} {{\zeta ^{({D_i} - {D_{i + u}})}}({\zeta ^{{s_i}}} + {\zeta ^{ - {s_{i + u}}}}) }}\\
{{ \times [1 + {{( - 1)}^{{i_{\pi (m - 1)}} - {{(i + u)}_{\pi (m - 1)}}}}]}  = 0}.
\end{array}
\end{equation}

\emph{Proof}: Firstly, we consider $u > 0$, let $k = i + u$ have a binary representation $\underline{k} = (k_0, k_1,...,k_{m-1})$.

\emph{Case 1.1}: $i_{\pi(m-1)} \neq k_{\pi(m-1)}$. Obviously, ${{\zeta ^{{D_i} - {D_{k}}}} \times [1 + {{( - 1)}^{{i_{\pi (m - 1)}} - {{k}_{\pi (m - 1)}}}}][{\zeta ^{ - s_{k}}} + {\zeta ^{s_i}}]} = 0$.

\emph{Case 1.2}: $i_{\pi(m-1)} = k_{\pi(m-1)}$, \eqref{eq lemma1} can be rewritten as $2 \times \sum\limits_{u = 1}^{n - 1} {\sum\limits_{i = 0}^{n - u - 1} {{\zeta ^{({D_i} - {D_{k}})}}({\zeta ^{{s_i}}} + {\zeta ^{ - {s_{k}}}})}  = 0} $. Let $v$ denote the biggest index for which $i_{\pi(v)} \neq k_{\pi(v)}$, where $1 \leq v \leq m-2$. Let $i'$ and $k'$ be the integers whose binary representations differ from those of $i$ and $k$ only at position $\pi(v+1)$, respectively, i.e.,
\begin{equation}
i' = ({i_0},{i_1},...,1 - {i_{\pi (v + 1)}},...,{i_{\pi (m - 1)}}),\nonumber
\end{equation}
\begin{equation}
k' = ({k_0},{k_1},...,1 - {k_{\pi (v + 1)}},...,{k_{\pi (m - 1)}}).\nonumber
\end{equation}

Due to $i_{\pi (v + 1)} = k_{\pi (v + 1)}$, we have $k' = i' + u$. Therefore, we define an invertible mapping $(i,k) \rightarrow (i',k')$.

According to \eqref{eq_qam16_ap1} and the definition of $v$, we have
\begin{equation}
\begin{array}{l}
{D_i} - {D_k} = 2\sum\limits_{l = 0}^{m - 2} {[{i_{\pi (l)}}{i_{\pi (l + 1)}} - {k_{\pi (l)}}{k_{\pi (l + 1)}}] }\\
{\rm{~~~~~~~~~~~~~~~~~~~~~~~}}{+ \sum\limits_{l = 0}^v {{c_{\pi (l)}}({i_{\pi (l)}} - {k_{\pi (l)}})} } ,\nonumber
\end{array}
\end{equation}
\begin{equation}
\begin{array}{l}
{D_{i'}} - {D_{k'}} = 2\sum\limits_{l = 0}^{m - 2} {[{i'_{\pi (l)}}{i'_{\pi (l + 1)}} - {k'_{\pi (l)}}{k'_{\pi (l + 1)}}] }\\
{\rm{~~~~~~~~~~~~~~~~~~~~~~~}}{+ \sum\limits_{l = 0}^v {{c_{\pi (l)}}({i'_{\pi (l)}} - {k'_{\pi (l)}})} } ,\nonumber
\end{array}
\end{equation}
\begin{equation}
\begin{array}{l}
{\rm{~~}}({D_{i}} - {D_{k}}) - ({D_{i'}} - {D_{k'}}) \\
= \sum\limits_{l = 0}^{m - 2} {[{i_{\pi (l)}}{i_{\pi (l + 1)}} - i{'_{\pi (l)}}i{'_{\pi (l + 1)}}]  } \\
{\rm{~~}}-\sum\limits_{l = 0}^{m - 2} {[{k_{\pi (l)}}{k_{\pi (l + 1)}} - k{'_{\pi (l)}}k{'_{\pi (l + 1)}}]  } \\
{\rm{~~}}+ \sum\limits_{l = 0}^v {{c_{\pi (l)}}({i_{\pi (l)}} - i{'_{\pi (l)}})}  - \sum\limits_{l = 0}^v {{c_{\pi (l)}}({k_{\pi (l)}} - k{'_{\pi (l)}})} \\
= 2[{i_{\pi (v)}}{i_{\pi (v + 1)}} - i{'_{\pi (v)}}i{'_{\pi (v + 1)}}] \\
{\rm{~~}}- 2[{k_{\pi (v)}}{k_{\pi (v + 1)}} - k{'_{\pi (v)}}k{'_{\pi (v + 1)}}] \\
= 2{i_{\pi (v)}} - 2{k_{\pi (v)}} = 2. \nonumber
\end{array}
\end{equation}

Therefore,
\begin{equation}
{\zeta ^{{D_{i'}} - {D_{k'}}}} =  - {\zeta ^{{D_i} - {D_k}}}. \nonumber
\end{equation}

If $v \geq 1$, according to the definition of $v$, we have ${s_i} = {s_{i'}}$ and ${s_k} = {s_{k'}}$. Hence,
\begin{equation}
{\zeta ^{{s_i}}} + {\zeta ^{ - {s_k}}} = {\zeta ^{{s_{i'}}}} + {\zeta ^{ - {s_{k'}}}}, \nonumber
\end{equation}
\begin{equation}
{\zeta ^{{D_{i'}} - {D_{k'}}}}({\zeta ^{{s_{i'}}}} + {\zeta ^{ - {s_{k'}}}}) + {\zeta ^{{D_i} - {D_k}}}({\zeta ^{{s_i}}} + {\zeta ^{ - {s_k}}}) = 0.\nonumber
\end{equation}

If $v = 0$, we have ${i'_{\pi(0)}} = {i_{\pi(0)}} \neq {k_{\pi(0)}} = {k'_{\pi(0)}}$ and ${i_{\pi(1)}} = {k_{\pi(1)}} \neq {i'_{\pi(1)}} = {k'_{\pi(1)}}$.

$(1)$ ${i_{\pi(1)}} = {k_{\pi(1)}} = 0 \Rightarrow {i'_{\pi(1)}} = {k'_{\pi(1)}} = 1$. And ${s_i} = {d_1}{i_{\pi (0)}} + {d_3}, {s_k} = {d_1}{k_{\pi (0)}} + {d_3}$, and ${s_{i'}} = 2{i_{\pi (0)}} + {d_1}{i_{\pi (0)}} + {d_2} + {d_3}, {s_{k'}} = 2{k_{\pi (0)}} + {d_1}{k_{\pi (0)}} + {d_2} + {d_3}$. According to the definition of \emph{Theorem 1}, we have
\begin{equation}
{s_i}+{s_k} = {d_1} + 2{d_3} = 2, \nonumber
\end{equation}
\begin{equation}
{s_{i'}}+{s_{k'}} = 2 + {d_1} + 2{d_2} + 2{d_3} = 2, \nonumber
\end{equation}
\begin{equation}
{\zeta ^{{s_i}}} + {\zeta ^{ - {s_k}}} = {\zeta ^{{s_{i'}}}} + {\zeta ^{ - {s_{k'}}}} = 0. \nonumber
\end{equation}
\begin{equation}
{\zeta ^{{D_{i'}} - {D_{k'}}}}({\zeta ^{{s_{i'}}}} + {\zeta ^{ - {s_{k'}}}}) + {\zeta ^{{D_i} - {D_k}}}({\zeta ^{{s_i}}} + {\zeta ^{ - {s_k}}}) = 0.\nonumber
\end{equation}

$(2)$ ${i_{\pi(1)}} = {k_{\pi(1)}} = 1 \Rightarrow {i'_{\pi(1)}} = {k'_{\pi(1)}} = 0$. Similarly,
\begin{equation}
{\zeta ^{{s_i}}} + {\zeta ^{ - {s_k}}} = {\zeta ^{{s_{i'}}}} + {\zeta ^{ - {s_{k'}}}} = 0, \nonumber
\end{equation}
\begin{equation}
{\zeta ^{{D_{i'}} - {D_{k'}}}}({\zeta ^{{s_{i'}}}} + {\zeta ^{ - {s_{k'}}}}) + {\zeta ^{{D_i} - {D_k}}}({\zeta ^{{s_i}}} + {\zeta ^{ - {s_k}}}) = 0.\nonumber
\end{equation}

Then, considering the situation of $u = 0$, \eqref{eq lemma1} can be rewritten as $2 \times \sum\limits_{i = 0}^{n - 1} {2 \times ({\zeta ^{{s_i}}} + {\zeta ^{ - {s_i}}})}=0$. Let $i'$ be the integer whose binary representation is $(1-{i_{\pi (0)}},{i_{\pi (1)}},...,{i_{\pi{ (m-1)}}})$, now an invertible mapping $i \rightarrow i'$ is defined, and $\sum\limits_{i = 0}^{n - 1} {({\zeta ^{{s_i}}} + {\zeta ^{ - {s_i}}})} {\rm{ = }}\sum\limits_{i = 0}^{n - 1} {({\zeta ^{{s_{i'}}}} + {\zeta ^{ - {s_{i'}}}})} $. According to \emph{Theorem 1},
\begin{equation}
{s_i} + {s_{i'}} = 2, \nonumber
\end{equation}
whenever ${i_{\pi(1)}} = 1$ or ${i_{\pi(1)}} = 0$. Therefore,
\begin{equation}
{\zeta ^{{s_i}}} + {\zeta ^{ - {s_i}}} + {\zeta ^{{s_{i'}}}} + {\zeta ^{ - {s_{i'}}}} = 0. \nonumber
\end{equation}

Since we have the mappings $(i,k) \rightarrow (i',k')$ and $i \rightarrow i'$ is invertible, the term  is equal to zero in \emph{Case 1.1}, and it sums to zero in pairs in \emph{Case 1.2}. Then we can have
\begin{equation}
\begin{array}{l}
 \sum\limits_{u = 1-n}^{n - 1} {\sum\limits_{i = 0}^{n - u - 1} {{\zeta ^{({D_i} - {D_{i + u}})}}({\zeta ^{{s_i}}} + {\zeta ^{ - {s_{i + u}}}})} } \\
  \times [1 + {( - 1)^{{i_{\pi (m - 1)}} - {{(i + u)}_{\pi (m - 1)}}}}] = 0. \nonumber
\end{array}
\end{equation}

Therefore, the proof of \emph{Lemma 1} is complete.

Now we are ready to give the proof of the \emph{Theorem 1}.

\emph{Proof of Theorem 1}: According to \eqref{eq Add16}, we can obtain that
\begin{equation}\label{eq_Ni_ap1}
\begin{split}
{\rm{H}} \star {{{\rm{H}}^{'}}} = &\sum\limits_{u = 1-n}^{n - 1} {\left| {C_{\rm{H}}}(u) + {C_{{{\rm{H}}^{'}}}}(u) \right|} \\
=& \sum\limits_{u = 0}^{n - 1} {\left| {\sum\limits_{i = 0}^{n - u - 1} {\{ r_1^2[{\zeta ^{{D_i} - {D_{i + u}}}} + {\rm{ }}{\zeta ^{{D_i}' - {D_{i + u}}'}}]\} } } \right.} \\
&  {+ r_2^2[{\zeta ^{{E_i} - {E_{i + u}}}} + {\zeta ^{E_i^{'} - E_{i + u}^{'}}}]  }  \\
&{ + {r_1}{r_2}{\zeta ^{{D_i} - {D_{i + u}}}}({\zeta ^{{s_i}}} + {\zeta ^{ - {s_{i + u}}}})} \\
&\left. { \times [1 + {{( - 1)}^{{i_{\pi (m - 1)}} - {{(i + u)}_{\pi (m - 1)}}}}]\} } \right|\\
\leq  &\sum\limits_{u = 1-n}^{n - 1} {\sum\limits_{i = 0}^{n - u - 1} {\left| {r_1^2[{\zeta ^{{D_i} - {D_{i + u}}}} + {\zeta ^{D_i^{'} - D_{i + u}^{'}}}]} \right|} } \\
&+ \sum\limits_{u = 1-n}^{n - 1} {\sum\limits_{i = 0}^{n - u - 1} {\left| {r_2^2[{\zeta ^{{E_i} - {E_{i + u}}}} + {\zeta ^{E_i^{'} - E_{i + u}^{'}}}]} \right|} } \\
&+{r_1}{r_2} \times \sum\limits_{u = 1-n}^{n - 1} {\sum\limits_{i = 0}^{n - u - 1} {\left| {{\zeta ^{{D_i} - {D_{i + u}}}}({\zeta ^{{s_i}}} + {\zeta ^{ - {s_{i + u}}}})} \right.} } \\
&\left. { \times [1 + {{( - 1)}^{{i_{\pi (m - 1)}} - {{(i + u)}_{\pi (m - 1)}}}}]} \right| .
\end{split}
\end{equation}

According to \emph{Lemma 1}, the term ${r_1}{r_2} \times \sum\limits_{u = 1-n}^{n - 1} {\sum\limits_{i = 0}^{n - u - 1} {\left| {{\zeta ^{{D_i} - {D_{i + u}}}}({\zeta ^{{s_i}}} + {\zeta ^{ - {s_{i + u}}}})} \right.} } \\ \left. { \times [1 + {{( - 1)}^{{i_{\pi (m - 1)}} - {{(i + u)}_{\pi (m - 1)}}}}]} \right|=0$, thus, \eqref{eq_Ni_ap1} can be rewritten as
\begin{equation}\label{eq_Ni_ap1_2}
\begin{split}
{\rm{H}} \star {{{\rm{H}}^{'}}} &\overset{\emph{Lemma 1}}{\leq}\sum\limits_{u = 1-n}^{n - 1} {\sum\limits_{i = 0}^{n - u - 1} {\left| {r_1^2[{\zeta ^{{D_i} - {D_{i + u}}}} + {\zeta ^{D_i^{'} - D_{i + u}^{'}}}]} \right|} } \\
 &{\rm{~~~~~~~~}}+ \sum\limits_{u = 1-n}^{n - 1} {\sum\limits_{i = 0}^{n - u - 1} {\left| {r_2^2[{\zeta ^{{E_i} - {E_{i + u}}}} + {\zeta ^{E_i^{'} - E_{i + u}^{'}}}]} \right|} } \\
&\overset{\eqref{eq 17a}}{=}r_1^2 D \star {D^{'}} + r_2^2 E \star {E^{'}}.
\end{split}
\end{equation}

According to \cite{Davis99}, it is easy to verify that the sequence pair $(D,{D^{'}})$ is a Golay complementary pair, and the sequence pair $(E,{E^{'}})$ is a near-complementary pair. Thus, the PMEPR of sequence $\rm{D}$ is at most $2$ (according to \eqref{eq pmepr3}), while the PMEPR of sequence $\rm{E}$ is at most $4$ (Theorem $10$ in \cite{Schmidt06} and Corollary $1$ in \cite{Lee10}), i.e.,
\begin{equation}\label{eq p1}
\begin{split}
D \star {D^{'}} \leq 2n,\\
E \star {E^{'}} \leq 4n.
\end{split}
\end{equation}

Substituting \eqref{eq p1} into \eqref{eq_Ni_ap1_2}, we can obtain that
\begin{equation}\label{eq_Ni_ap1_3}
\begin{split}
{\rm{H}} \star {{{\rm{H}}^{'}}} &\leq 2r_1^2 n + 4r_2^2 n\\
&=0.8 \times 2n + 0.2 \times 4n\\
& = 2.4n \ll 2n^2.
\end{split}
\end{equation}

Therefore, the proposed $16$-QAM sequence $\textrm{H}$ is a near-complementary sequence because $2n \le \textrm{H} \star \textrm{H}^{'} \ll 2{n^2}$.

Moreover, according to \eqref{eq pmepr2} and \eqref{eq_Ni_ap1_3}, the PMEPR of the sequence $\textrm{H}$ is
\begin{equation}
\begin{split}
{\rm{PMEPR}}(\textrm{H}) \leq (\textrm{H} \star \textrm{H}^{'})/n \le 2.4.
\end{split}
\end{equation}

Therefore, the proof of \emph{Theorem 1} is complete.

\begin{center}
APPENDIX \uppercase\expandafter{\romannumeral2}
\end{center}
\begin{center}
The Proof of \emph{Theorem 2}
\end{center}

Firstly, we recall \eqref{eq s3} and \eqref{eq_qam64}, i.e.,

\begin{equation}\label{eq s3_ap2}
\begin{split}
D(\underline{x}) &= 2\sum\limits_{l = 0}^{m - 2} {{x_{\pi (l)}}{x_{\pi (l + 1)}} + \sum\limits_{l = 0}^{m - 1} {{c_l}{x_{\pi (l)}} + c}}.\\
F(\underline{x}) &= D(\underline{x}) + s^{(1)}(\underline{x}),\\
G(\underline{x}) &= D(\underline{x}) + s^{(2)}(\underline{x}),\\
{D^{'}}(\underline{x}) &= D(\underline{x}) + 2{x_{\pi (m - 1)}}, \\
{F^{'}}(\underline{x}) &= F(\underline{x}) + 2{x_{\pi (m - 1)}}, \\
{G^{'}}(\underline{x}) &= G(\underline{x}) + 2{x_{\pi (m - 1)}}, \\
{\textrm{J}(\underline{x})} &= \gamma ({a_1}{\zeta ^{{D(\underline{x})}}} + {a_2}{\zeta ^{{F(\underline{x})}}} + {a_3}{\zeta ^{{G (\underline{x})}}}).
\end{split}
\end{equation}
Then, we define ${s^{(3)}}(\underline{x})$ and ${\textrm{J}^{'}}(\underline{x})$ as follows
\begin{equation}\label{eq s3_ap2_2}
\begin{split}
{s^{(3)}}(\underline{x}) &= {s^{(1)}}(\underline{x}) - {s^{(2)}}(\underline{x}), \\
{\textrm{J}^{'}}(\underline{x}) &= \gamma ({a_1}{\zeta ^{{{D^{'}}(\underline{x})}}} + {a_2}{\zeta ^{{{F^{'}}(\underline{x})}}} + {a_3}{\zeta ^{{{G^{'}}(\underline{x})}}}).
\end{split}
\end{equation}

According to \eqref{eq 16a} and \eqref{eq s3_ap2}, the aperiodic auto-correlation function of $\textrm{J}(\underline{x})$ can be expressed as follows
\begin{equation}\label{eq_1}
\begin{array}{l}
{\rm{~~}}{C_{\textrm{J}}}(u) \\
= \sum\limits_{i = 1}^{n - u - 1} {({a_1}{\zeta ^{{D_i}}} + {a_2}{\zeta ^{{F_i}}} + {a_3}{\zeta ^{{G_i}}})}  \\
 \times {({a_1}{\zeta ^{{D_{i+u}}}} + {a_2}{\zeta ^{{F_{i+u}}}} + {a_3}{\zeta ^{{G_{i+u}}}})^*} \\
= \sum\limits_{i = 0}^{n - u - 1} {[a_1^2{\zeta ^{{D_i} - {D_{i + u}}}} + a_2^2} {\zeta ^{{F_i} - {F_{i + u}}}} + a_3^2{\zeta ^{{G_i} - {G_{i + u}}}}  \\
+ {a_1}{a_2}({\zeta ^{{D_i} - {F_{i + u}}}} + {\zeta ^{{F_i} - {D_{i + u}}}}) \\
+ {a_1}{a_3}({\zeta ^{{D_i} - {G_{i + u}}}} + {\zeta ^{{G_i} - {D_{i + u}}}}) \\
+ {a_2}{a_3}({\zeta ^{{F_i} - {G_{i + u}}}} + {\zeta ^{{G_i} - {F_{i + u}}}})].
\end{array}
\end{equation}

Similarly, according to \eqref{eq 16a} and \eqref{eq s3_ap2_2} the aperiodic auto-correlation function of $\textrm{J}(\underline{x})^{'}$ can be expressed as
\begin{equation}\label{eq_1_2}
\begin{array}{l}
{\rm{~~}}{C_{\textrm{J}^{'}}}(u) \\
= \sum\limits_{i = 1}^{n - u - 1} {({a_1}{\zeta ^{{{D^{'}}_i}}} + {a_2}{\zeta ^{{{F^{'}}_i}}} + {a_3}{\zeta ^{{{G^{'}} _i}}})} \\
 \times {({a_1}{\zeta ^{{{D^{'}}_{i+u}}}} + {a_2}{\zeta ^{{{F^{'}}_{i+u}}}} + {a_3}{\zeta ^{{{G^{'}}_{i+u}}}})^ * } \\
= \sum\limits_{i = 0}^{n - u - 1} {[a_1^2{\zeta ^{{D_i^{'}} - {D_{i + u}^{'}}}} + a_2^2} {\zeta ^{{F_i^{'}} - {F_{i + u}^{'}}}} + a_3^2{\zeta ^{{G_i^{'}} - {G_{i + u}^{'}}}} \\
+ {a_1}{a_2}({\zeta ^{{G_i^{'}} - {F_{i + u}^{'}}}} + {\zeta ^{{F_i} - {D_{i + u}^{'}}}}) \\
+ {a_1}{a_3}({\zeta ^{{D_i^{'}} - {G _{i + u}^{'}}}} + {\zeta ^{{G_i^{'}} - {D_{i + u}^{'}}}}) \\
+ {a_2}{a_3}({\zeta ^{{F_i^{'}} - {G_{i + u}^{'}}}} + {\zeta ^{{G_i^{'}} - {F_{i + u}^{'}}}})]. \\
\end{array}
\end{equation}

Then, combine \eqref{eq_1} and \eqref{eq_1_2}, we have
\begin{equation}\label{eq_Ni_ap2_5}
\begin{array}{l}
{\rm{~~}}{C_{\textrm{J}}}(u) + {C_{\textrm{J}^{'}}}(u) \\
= \sum\limits_{i = 0}^{n - u - 1} {[a_1^2({\zeta ^{{D_i} - {D_{i + u}}}} + {\zeta ^{{D_i^{'}} - {D_{i + u}^{'}}}})}  \\
+ a_2^2({\zeta ^{{F_i} - {F_{i + u}}}} + {\zeta ^{{F_i^{'}} - {F_{i + u}^{'}}}}) \\
+ a_3^2({\zeta ^{{G_i} - {G_{i + u}}}} + {\zeta ^{{G_i^{'}} - {G_{i + u}^{'}}}}) \\
+ {a_1}{a_2}({\zeta ^{{D_i} - {F_{i + u}}}} + {\zeta ^{{F_i} - {D_{i + u}}}} + {\zeta ^{{D_i^{'}} - {F_{i + u}^{'}}}} + {\zeta ^{{F_i^{'}} - {D_{i + u}^{'}}}}) \\
+ {a_1}{a_3}({\zeta ^{{D_i} - {G_{i + u}}}} + {\zeta ^{{G_i} - {D_{i + u}}}} + {\zeta ^{{D_i^{'}} - {G_{i + u}^{'}}}} + {\zeta ^{{G_i^{'}} - {D_{i + u}^{'}}}})  \\
+{a_2}{a_3}({\zeta ^{{F_i} - {G_{i + u}}}} + {\zeta ^{{G_i} - {F_{i + u}}}} + {\zeta ^{{F_i^{'}} - {G_{i + u}^{'}}}} + {\zeta ^{{G_i^{'}} - {F_{i + u}^{'}}}})].
\end{array}
\end{equation}

The last three terms in \eqref{eq_Ni_ap2_5} can be rewritten as follows
\begin{equation}\label{eq_Ni_ap2_6}
\begin{array}{l}
{\rm{~~~}}{a_1}{a_2}({\zeta ^{{D_i} - {F_{i + u}}}} + {\zeta ^{{F_i} - {D_{i + u}}}} + {\zeta ^{{D_i^{'}} - {F_{i + u}^{'}}}} + {\zeta ^{{F_i^{'}} - {D_{i + u}^{'}}}})  \\
 = {a_1}{a_2}\sum\limits_{i = 0}^{n - u - 1} {{\zeta ^{{D_i} - {D_{i + u}}}}} \\
 {\rm{~~~~~~~~~~~~~~}} { \times [1 + {{( - 1)}^{{i_{\pi (m - 1)}} - {{(i + u)}_{\pi (m - 1)}}}}][{\zeta ^{ - s_{i + u}^{(1)}}} + {\zeta ^{s_i^{(1)}}}]},
\end{array}
\end{equation}
\begin{equation}
\begin{array}{l}
{\rm{~~~}}{a_1}{a_3}({\zeta ^{{D_i} - {G_{i + u}}}} + {\zeta ^{{G_i} - {D_{i + u}}}} + {\zeta ^{{D_i^{'}} - {G_{i + u}^{'}}}} + {\zeta ^{{G_i^{'}} - {D_{i + u}^{'}}}})   \\
= {a_1}{a_3}\sum\limits_{i = 0}^{n - u - 1} {{\zeta ^{{D_i} - {D_{i + u}}}} }\\
   {\rm{~~~~~~~~~~~~~~}} {\times [1 + {{( - 1)}^{{i_{\pi(m - 1)}} - {{(i + u)}_{\pi(m - 1)}}}}][{\zeta ^{ - s_{i + u}^{(2)}}} + {\zeta ^{s_i^{(2)}}}]},
\end{array}
\end{equation}
\begin{equation}
\begin{array}{l}
{\rm{~~~}}{a_2}{a_3}({\zeta ^{{F_i} - {G_{i + u}}}} + {\zeta ^{{G_i} - {F_{i + u}}}} + {\zeta ^{{F_i^{'}} - {G_{i + u}^{'}}}} + {\zeta ^{{G_i^{'}} - {F_{i + u}^{'}}}})   \\
= {a_2}{a_3}\sum\limits_{i = 0}^{n - u - 1} {{\zeta ^{{F_i} - {F_{i + u}}}} }\\
  {\rm{~~~~~~~~~~~~~~}} {\times [1 + {{( - 1)}^{{i_{\pi (m - 1)}} - {{(i + u)}_{\pi (m - 1)}}}}][{\zeta ^{ - s_{i + u}^{(3)}}} + {\zeta ^{s_i^{(3)}}}]}.
\end{array}
\end{equation}

To prove \emph{Theorem 2}, we should firstly prove \emph{Lemma 2} and \emph{Lemma 3}.

\emph{Lemma 2}: In \emph{Type 1}, the following relationships can be obtained
\begin{equation}\label{r1r2}
\begin{array}{l}
 {a_1}{a_2}\sum\limits_{u = 1}^{n - 1}\bigg|{\sum\limits_{i = 0}^{n - u - 1} {{\zeta ^{{D_i} - {D_{i + u}}}}}  } \\
  \times [1 + {{( - 1)}^{{i_{\pi (m - 1)}} - {{(i + u)}_{\pi (m - 1)}}}}]\times ({\zeta ^{ - s_{i + u}^{(1)}}} + {\zeta ^{s_i^{(1)}}})\bigg| = 0,
 \end{array}
\end{equation}
\begin{equation}\label{r1r3}
\begin{array}{l}
{a_1}{a_3}\sum\limits_{u = 1-n}^{n - 1} \bigg|{\sum\limits_{i = 0}^{n - u - 1} {{\zeta ^{{D_i} - {D_{i + u}}}}} } \\
 \times [1 + {{( - 1)}^{{i_{\pi (m - 1)}} - {{(i + u)}_{\pi (m - 1)}}}}] \times ({\zeta ^{ - s_{i + u}^{(2)}}} + {\zeta ^{s_i^{(2)}}}) \bigg| = 0,
 \end{array}
\end{equation}
\begin{equation}\label{r2r3}
\begin{array}{l}
{a_2}{a_3}\sum\limits_{u = 1-n}^{n - 1} \bigg|{\sum\limits_{i = 0}^{n - u - 1} {{\zeta ^{{F_i} - {F_{i + u}}}}} }\\
  \times [1 + {{( - 1)}^{{i_{\pi (m - 1)}} - {{(i + u)}_{\pi (m - 1)}}}}] \times ({\zeta ^{ - s_{i + u}^{(3)}}} + {\zeta ^{s_i^{(3)}}})\bigg| = 0.
 \end{array}
\end{equation}

\emph{Proof}: According to \emph{Lemma 1}, it is obvious that \eqref{r1r3} is satisfied. The offset ${s^{(1)}}(\underline{x})$ is in subset cases of \cite{Chong03}. Based on the proof in \cite{Chong03} and \cite{Li08}, we can prove that \eqref{r1r2} is satisfied. Then we will prove \eqref{r2r3}.

Firstly, we consider $u > 0$, let $k = i + u$ have a binary representation $\underline{k} = (k_0, k_1,...,k_{m-1})$.

\emph{Case 2.1}: $i_{\pi(m-1)} \neq k_{\pi(m-1)}$. Obviously, ${{\zeta ^{{F_i} - {F_{k}}}} \times [1 + {{( - 1)}^{{i_{\pi (m - 1)}} - {{k}_{\pi (m - 1)}}}}]\times({\zeta ^{ - s_{k}^{(3)}}} + {\zeta ^{s_i^{(3)}}})} = 0$.

\emph{Case 2.2}: $i_{\pi(m-1)} = k_{\pi(m-1)}$, \eqref{r2r3} can be rewritten $ {a_2}{a_3} \sum\limits_{u = 1}^{n - 1}\bigg|2 \times {\sum\limits_{i = 0}^{n - u - 1} {{\zeta ^{({F_i} - {F_{k}})}}({\zeta ^{{s_i^{(3)}}}} + {\zeta ^{ - {s_{k}^{(3)}}}})} \bigg| = 0} $. Let $v$ denote the biggest index for which $i_{\pi(v)} \neq k_{\pi(v)}$, where $v \leq m-2 $. Denote $i'$ and $k'$ as the integers whose binary representations differ from those of $i$ and $k$ only at position $\pi(v+1)$, respectively, i.e.,
\begin{equation}
\begin{split}
i' = ({i_0},{i_1},...,1 - {i_{\pi (v + 1)}},...,{i_{\pi (m - 1)}}),\\
k' = ({k_0},{k_1},...,1 - {k_{\pi (v + 1)}},...,{k_{\pi (m - 1)}}),\\
\end{split}
\end{equation}
since $i_{\pi (v + 1)} = k_{\pi (v + 1)}$, we can obtain $k' = i' + u$. Then, we define an invertible mapping $(i,k) \rightarrow (i',k')$.

In \emph{Type 1}, we have
\begin{equation}
\begin{array}{l}
  {\rm{~~~}}{F_i} - {F_k}\\
 = 2\sum\limits_{l = 0}^{m - 2} {[{i_{\pi (l)}}{i_{\pi (l + 1)}} - {k_{\pi (l)}}{k_{\pi (l + 1)}}] } \\
 {\rm{~~~}}{+ \sum\limits_{l = 0}^v {{c_{\pi (l)}}({i_{\pi (l)}} - {k_{\pi (l)}})} } + s_i^{(1)} - s_k^{(1)} ,\\
\end{array}
\end{equation}
\begin{equation}
\begin{array}{l}
{\rm{~~~}}{F_{i'}} - {F_{k'}} \\
= 2\sum\limits_{l = 0}^{m - 2} {[{i'_{\pi (l)}}{i'_{\pi (l + 1)}} - {k'_{\pi (l)}}{k'_{\pi (l + 1)}}] } \\
{\rm{~~~}}{+ \sum\limits_{l = 0}^v {{c_{\pi (l)}}({i'_{\pi (l)}} - {k'_{\pi (l)}})} } + s_{i'}^{(1)} - s_{k'}^{(1)},\\
\end{array}
\end{equation}
\begin{equation}
\begin{array}{l}
{\rm{~~~}}({F_{i}} - {F_{k}}) - ({F_{i'}} - {F_{k'}}) \\
= 2[{i_{\pi (v)}}{i_{\pi (v + 1)}} - i{'_{\pi (v)}}i{'_{\pi (v + 1)}}] \\
{\rm{~~~}}- 2[{k_{\pi (l)}}{k_{\pi (l + 1)}} - k{'_{\pi (l)}}k{'_{\pi (l + 1)}}]\\
{\rm{~~~}} + s_i^{(1)} - s_k^{(1)} - s_{i'}^{(1)} + s_{k'}^{(1)} \\
= 2 + s_i^{(1)} - s_k^{(1)} - s_{i'}^{(1)} + s_{k'}^{(1)}. \\
\end{array}
\end{equation}

According to \cite{Chong03} and \cite{Li08}, we have $s_i^{(1)} - s_k^{(1)} - s_{i'}^{(1)} + s_{k'}^{(1)} = 0$ for all $u > 0$. Therefore,
\begin{equation}
{\zeta ^{{F_{i'}} - {F_{k'}}}} =  - {\zeta ^{{F_i} - {F_k}}}.
\end{equation}

If $v \geq 1$, with the definition of $v$, we obtain $s_i^{(3)} = s_{i'}^{(3)}$ and $s_k^{(3)} = s_{k'}^{(3)}$. Hence,
\begin{equation}
{\zeta ^{s_i^{(3)}}} + {\zeta ^{ - s_k^{(3)}}} = {\zeta ^{s_{i'}^{(3)}}} + {\zeta ^{ - s_{k'}^{(3)}}},
\end{equation}
\begin{equation}
{\zeta ^{{F_{i'}} - {F_{k'}}}}({\zeta ^{s_{i'}^{(3)}}} + {\zeta ^{ - s_{k'}^{(3)}}}) + {\zeta ^{{F_i} - {F_k}}}({\zeta ^{s_i^{(3)}}} + {\zeta ^{ - s_k^{(3)}}}) = 0.
\end{equation}

If $v = 0$, we have ${i'_{\pi(0)}} = {i_{\pi(0)}} \neq {k_{\pi(0)}} = {k'_{\pi(0)}}$ and ${i_{\pi(1)}} = {k_{\pi(1)}} \neq {i'_{\pi(1)}} = {k'_{\pi(1)}}$.

$(1)$ ${i_{\pi(1)}} = {k_{\pi(1)}} = 1 \Rightarrow {i'_{\pi(1)}} = {k'_{\pi(1)}} = 0$. So $s_i^{(3)} = ({h_1} - {d_1}){i_{\pi (0)}} + {h_3} - {d_3}, {s_k^{(3)}} = ({h_1} - {d_1}){k_{\pi (0)}} + {h_3} - {d_3}$, and $s_{i'}^{(3)} = 2{i_{\pi (0)}}{i_{\pi (1)}} + ({h_1} - {d_1}){i_{\pi (0)}}  - {d_2} + {h_3} - {d_3}, s_{k'}^{(3)} = 2{k_{\pi (0)}}{k_{\pi (1)}} + ({h_1} - {d_1}){k_{\pi (0)}} - {d_2} + {h_3} - {d_3}$. Moreover,
\begin{equation}
{s_i^{(3)}}+{s_k^{(3)}} = {h_1} + 2{h_3} - {d_1} - 2{d_3} = 2,
\end{equation}
\begin{equation}
{s_{i'}^{(3)}}+{s_{k'}^{(3)}} = 2 + {h_1} + 2{h_3} - {d_1} - 2{d_2} - 2{d_3} = 2,
\end{equation}
\begin{equation}
{\zeta ^{{s_i^{(3)}}}} + {\zeta ^{ - {s_k^{(3)}}}} = {\zeta ^{{s_{i'}^{(3)}}}} + {\zeta ^{ - {s_{k'}^{(3)}}}} = 0,
\end{equation}
\begin{equation}
{\zeta ^{{F_{i'}} - {F_{k'}}}}({\zeta ^{s_{i'}^{(3)}}} + {\zeta ^{ - s_{k'}^{(3)}}}) + {\zeta ^{{F_i} - {F_k}}}({\zeta ^{s_i^{(3)}}} + {\zeta ^{ - s_k^{(3)}}}) = 0.
\end{equation}

$(2)$ ${i_{\pi(1)}} = {k_{\pi(1)}} = 0 \Rightarrow {i'_{\pi(1)}} = {k'_{\pi(1)}} = 1$. Similarly, we have
\begin{equation}
{\zeta ^{s_i^{(3)}}} + {\zeta ^{ - s_k^{(3)}}} = {\zeta ^{s_{i'}^{(3)}}} + {\zeta ^{ - s_{k'}^{(3)}}} = 0,
\end{equation}
\begin{equation}
{\zeta ^{{F_{i'}} - {F_{k'}}}}({\zeta ^{s_{i'}^{(3)}}} + {\zeta ^{ - s_{k'}^{(3)}}}) + {\zeta ^{{F_i} - {F_k}}}({\zeta ^{s_i^{(3)}}} + {\zeta ^{ - s_k^{(3)}}}) = 0.
\end{equation}

Then, we consider the situation when $u = 0$. The term \eqref{r2r3} can be rewritten as $\bigg| 2 \times {a_2}{a_3}\sum\limits_{i = 0}^{n - 1} {({\zeta ^{{s_i^{(3)}}}} + {\zeta ^{ - {s_i^{(3)}}}})}\bigg| =0$. Let $i'$ be the integer whose binary representation is $(1-{i_{\pi (0)}},{i_{\pi (1)}},...,{i_{\pi{ (m-1)}}})$, an invertible mapping $i \rightarrow i'$ is defined, $\sum\limits_{i = 0}^{n - 1} {({\zeta ^{{s_i}}} + {\zeta ^{ - {s_i}}})} {\rm{ = }}\sum\limits_{i = 0}^{n - 1} {({\zeta ^{{s_{i'}}}} + {\zeta ^{ - {s_{i'}}}})} $. According to \emph{Type 1},
\begin{equation}
{s_i^{(3)}} + {s_{i'}^{(3)}} = 2,
\end{equation}
whenever ${i_{\pi(1)}} = 1$ or ${i_{\pi(1)}} = 0$. Therefore,
\begin{equation}\label{eq v000}
{\zeta ^{s_i^{(3)}}} + {\zeta ^{ - s_{i}^{(3)}}} + {\zeta ^{s_{i'}^{(3)}}} + {\zeta ^{ - s_{i'}^{(3)}}} = 0.
\end{equation}

Similar to \emph{Lemma 1}, \emph{Lemma 2} is proven to be true. The proof of \emph{Lemma 2} is complete.

\emph{Lemma 3}: In \emph{Type 2}, it can be obtained that
\begin{equation}\label{r1r2'}
\begin{array}{l}
 {a_1}{a_2}\sum\limits_{u = 1-n}^{n - 1} \bigg| {\sum\limits_{i = 0}^{n - u - 1} {{\zeta ^{{D_i} - {D_{i + u}}}} }} \\{{ \times [1 + {{( - 1)}^{{i_{\pi (m - 1)}} - {{(i + u)}_{\pi (m - 1)}}}}]\times({\zeta ^{ - s_{i + u}^{(1)}}} + {\zeta ^{s_i^{(1)}}})} \bigg| = 0},
 \end{array}
\end{equation}
\begin{equation}\label{r1r3'}
\begin{array}{l}
{a_1}{a_3}\sum\limits_{u = 1-n}^{n - 1} \bigg| {\sum\limits_{i = 0}^{n - u - 1} {{\zeta ^{{D_i} - {D_{i + u}}}}}}\\ {{\times [1 + {{( - 1)}^{{i_{\pi (m - 1)}} - {{(i + u)}_{\pi (m - 1)}}}}]\times({\zeta ^{ - s_{i + u}^{(2)}}} + {\zeta ^{s_i^{(2)}}})} \bigg| = 0},
\end{array}
\end{equation}
\begin{equation}\label{r2r3'}
\begin{array}{l}
{a_2}{a_3}\sum\limits_{u = 1-n}^{n - 1} \bigg| {\sum\limits_{i = 0}^{n - u - 1} {{\zeta ^{{F_i} - {F_{i + u}}}} }} \\{{\times [1 + {{( - 1)}^{{i_{\pi (m - 1)}} - {{(i + u)}_{\pi (m - 1)}}}}]\times({\zeta ^{ - s_{i + u}^{(3)}}} + {\zeta ^{s_i^{(3)}}})} \bigg| = 0}.
  \end{array}
\end{equation}

\emph{Proof}: Because the sequence pair $(D,{D^{'}})$ is a Golay complementary pair, and both $(F,{F^{'}})$ and $(G,{G^{'}})$ are near-complementary pairs in \emph{Case 2}. According to \emph{Lemma 1}, we can easily verify that \eqref{r1r2'} and \eqref{r1r3'} are satisfied. Now we just need to prove \eqref{r2r3'} is true.

Firstly, we consider $u > 0$, let $k = i + u$ have a binary representation $\underline{k} = (k_0, k_1,...,k_{m-1})$.

\emph{Case 3.1}: $i_{\pi(m-1)} \neq k_{\pi(m-1)}$. Obviously, ${{\zeta ^{{F_i} - {F_{k}}}} \times [1 + {{( - 1)}^{{i_{\pi (m - 1)}} - {{k}_{\pi (m - 1)}}}}]\times({\zeta ^{ - s_{k}^{(3)}}} + {\zeta ^{s_i^{(3)}}})} = 0$.

\emph{Case 3.2}: $i_{\pi(m-1)} = k_{\pi(m-1)}$, \eqref{r2r3'} can be rewritten as $  {a_2}{a_3} \sum\limits_{u = 0}^{n - 1}\bigg|2\times {\sum\limits_{i = 0}^{n - u - 1} {{\zeta ^{({F_i} - {F_{k}})}}({\zeta ^{{s_i^{(3)}}}} + {\zeta ^{ - {s_{k}^{(3)}}}})} \bigg| = 0} $. Let $v$ denote the biggest index for which $i_{\pi(v)} \neq k_{\pi(v)}$, where $v \leq m-2$. Let $i'$ and $k'$ denote indexes whose binary representations differ from those of $i$ and $k$ only at position $\pi(v+1)$, respectively, i.e.,
\begin{equation}
\begin{split}
i' = ({i_0},{i_1},...,1 - {i_{\pi (v + 1)}},...,{i_{\pi (m - 1)}}),\\
k' = ({k_0},{k_1},...,1 - {k_{\pi (v + 1)}},...,{k_{\pi (m - 1)}}),\\
\end{split}
\end{equation}
since $i_{\pi (v + 1)} = k_{\pi (v + 1)}$, we can obtain $k' = i' + u$. Then, we define an invertible mapping $(i,k) \rightarrow (i',k')$.

In \emph{Type 2}, we have
\begin{equation}
\begin{array}{l}
{\rm{~~~}}{F_i} - {F_k}\\
 = 2\sum\limits_{l = 0}^{m - 2} {[{i_{\pi (l)}}{i_{\pi (l + 1)}} - {k_{\pi (l)}}{k_{\pi (l + 1)}}] } \\
 {\rm{~~~}}{+ \sum\limits_{l = 0}^v {{c_{\pi (l)}}({i_{\pi (l)}} - {k_{\pi (l)}})} } + s_i^{(1)} - s_k^{(1)},
\end{array}
\end{equation}
\begin{equation}
\begin{array}{l}
 {\rm{~~~}}{F_{i'}} - {F_{k'}}\\
= 2\sum\limits_{l = 0}^{m - 2} {[{i'_{\pi (l)}}{i'_{\pi (l + 1)}} - {k'_{\pi (l)}}{k'_{\pi (l + 1)}}] }\\
  {\rm{~~~}}{+ \sum\limits_{l = 0}^v {{c_{\pi (l)}}({i'_{\pi (l)}} - {k'_{\pi (l)}})} } + s_{i'}^{(1)} - s_{k'}^{(1)},
\end{array}
\end{equation}
\begin{equation}
\begin{array}{l}
{\rm{~~}}({F_{i'}} - {F_{k'}}) - ({F_i} - {F_k}) \\
= 2[{i_{\pi (v)}}{i_{\pi (v + 1)}} - i{'_{\pi (v)}}i{'_{\pi (v + 1)}}] \\
- 2[{k_{\pi (l)}}{k_{\pi (l + 1)}} - k{'_{\pi (l)}}k{'_{\pi (l + 1)}}] \\
+ s_i^{(1)} - s_k^{(1)} - s_{i'}^{(1)} + s_{k'}^{(1)} \\
= 2 + s_i^{(1)} - s_k^{(1)} - s_{i'}^{(1)} + s_{k'}^{(1)}.
\end{array}
\end{equation}

If $v \geq 1$, with the definition of $v$, we have $s_i^{(1)} = s_{i'}^{(1)}$, $s_k^{(1)} = s_{k'}^{(1)}$, $s_i^{(3)} = s_{i'}^{(3)}$ and $s_k^3 = s_{k'}^3$.

Hence,
\begin{equation}
{\zeta ^{{F_{i'}} - {F_{k'}}}} =  - {\zeta ^{{F_i} - {F_k}}},
\end{equation}
\begin{equation}
{\zeta ^{s_i^{(3)}}} + {\zeta ^{ - s_k^{(3)}}} = {\zeta ^{s_{i'}^{(3)}}} + {\zeta ^{ - s_{k'}^{(3)}}},
\end{equation}
\begin{equation}
{\zeta ^{{F_{i'}} - {F_{k'}}}}({\zeta ^{s_{i'}^{(3)}}} + {\zeta ^{ - s_{k'}^{(3)}}}) + {\zeta ^{{F_i} - {F_k}}}({\zeta ^{s_i^{(3)}}} + {\zeta ^{ - s_k^{(3)}}}) = 0.
\end{equation}

If $v = 0$, we have ${i'_{\pi(0)}} = {i_{\pi(0)}} \neq {k_{\pi(0)}} = {k'_{\pi(0)}}$ and ${i_{\pi(1)}} = {k_{\pi(1)}} \neq {i'_{\pi(1)}} = {k'_{\pi(1)}}$.

$(1)$ ${i_{\pi(1)}} = {k_{\pi(1)}} = 1 \Rightarrow {i'_{\pi(1)}} = {k'_{\pi(1)}} = 0$. We have
\begin{equation}
s_i^{(1)} = {d_1}{i_{\pi (0)}} + {d_3}, s_i^{(3)} = ({d_1} - {h_1}){i_{\pi (0)}} + {d_3} - {h_3},
\end{equation}
\begin{equation}
s_k^{(1)} = {d_1}{j_{\pi (0)}} + {d_3}, s_k^{(3)} = ({d_1} - {h_1}){k_{\pi (0)}} + {d_3} - {h_3},
\end{equation}
\begin{equation}
\begin{split}
s_i^{(1)} + s_{k'}^{(1)} = 2{k_{\pi (0)}} + {d_1} + {d_2} + 2{d_3}, \\
s_k^{(1)} + s_{i'}^{(1)} = 2{k_{\pi (0)}} + {d_1} + {d_2} + 2{d_3},
\end{split}
\end{equation}
\begin{equation}
s_{i'}^{(3)} - s_i^{(3)} = 2, s_{k'}^{(3)} - s_k^{(3)} = 2,
\end{equation}
\begin{equation}
\begin{split}
s_{i'}^{(1)} - s_i^{(1)} + s_{k'}^{(1)} - s_j^{(1)} = 2 , \\
({F_{i'}} - {F_{k'}}) - ({F_i} - {F_k}) = 0.
\end{split}
\end{equation}

Therefore,
\begin{equation}
{\zeta ^{{F_i} - {F_k}}} = {\zeta ^{{F_{i'}} - {F_{k'}}}},
\end{equation}
\begin{equation}
{\zeta ^{s_i^{(3)}}} + {\zeta ^{ - s_k^{(3)}}} = - ({\zeta ^{s_{i'}^{(3)}}} + {\zeta ^{ - s_{k'}^{(3)}}}),
\end{equation}
\begin{equation}
{\zeta ^{{F_{i'}} - {F_{k'}}}}({\zeta ^{s_{i'}^{(3)}}} + {\zeta ^{ - s_{k'}^{(3)}}}) + {\zeta ^{{F_i} - {F_k}}}({\zeta ^{s_i^{(3)}}} + {\zeta ^{ - s_k^{(3)}}}) = 0.
\end{equation}

$(2)$ ${i_{\pi(1)}} = {k_{\pi(1)}} = 0 \Rightarrow {i'_{\pi(1)}} = {k'_{\pi(1)}} = 1$. Similarly, we have
\begin{equation}
{\zeta ^{{F_i} - {F_k}}} = {\zeta ^{{F_{i'}} - {F_{k'}}}},
\end{equation}
\begin{equation}
{\zeta ^{{s_i^{(3)}}}} + {\zeta ^{ - {s_k^{(3)}}}} = - ({\zeta ^{{s_{i'}^{(3)}}}} + {\zeta ^{ - {s_{k'}^{(3)}}}}),
\end{equation}
\begin{equation}
{\zeta ^{{F_{i'}} - {F_{k'}}}}({\zeta ^{s_{i'}^{(3)}}} + {\zeta ^{ - s_{k'}^{(3)}}}) + {\zeta ^{{F_i} - {F_k}}}({\zeta ^{s_i^{(3)}}} + {\zeta ^{ - s_k^{(3)}}}) = 0.
\end{equation}

Then, we consider the situation when $u = 0$. The term \eqref{r2r3'} can be rewritten as $\bigg| 2 \times {a_2}{a_3}\sum\limits_{i = 0}^{n - 1} { ({\zeta ^{{s_i^{(3)}}}} + {\zeta ^{ - {s_i^{(3)}}}})} \bigg| =0$. Let $i'$ be the integer whose binary representation is $({i_{\pi (0)}},1-{i_{\pi (1)}},...,{i_{\pi{ (m-1)}}})$, an invertible mapping $i \rightarrow i'$ is defined, $\sum\limits_{i = 0}^{n - 1} {({\zeta ^{{s_i^{(3)}}}} + {\zeta ^{ - {s_i^{(3)}}}})}  = \sum\limits_{i = 0}^{n - 1} {({\zeta ^{{s_{i'}^{(3)}}}} + {\zeta ^{ - {s_{i'}^{(3)}}}})}$. According to \emph{Type 2}, we have
\begin{equation}
{s_{i'}^{(3)}} - {s_i^{(3)}} = 2,
\end{equation}
whenever ${i_{\pi(1)}} = 1$ or ${i_{\pi(1)}} = 0$. Therefore,
\begin{equation}\label{eq v000}
{\zeta ^{s_i^{(3)}}} + {\zeta ^{ - s_i^{(3)}}} + {\zeta ^{s_{i'}^{(3)}}} + {\zeta ^{ - s_{i'}^{(3)}}} = 0.
\end{equation}

Obviously, similar to \emph{Lemma 2}, \emph{Lemma 3} is proven to be true. The proof of \emph{Lemma 3} is complete.

Now, the proof of \emph{Theorem 2} is given as follows.

\emph{Proof of Theorem 2}: According to \cite{Davis99}, Theorem $10$ in \cite{Schmidt06} and Corollary $1$ in \cite{Lee10}, for \emph{Type 1}, both the sequence pairs $(D,{D^{'}})$ and $(F,{F^{'}})$ are Golay complementary pairs, while $(G,{G^{'}})$ is a near-complementary pair. For \emph{Type 2}, both $(F,{F^{'}})$ and $(G,{G^{'}})$ are near-complementary pairs, while the sequence pair $(D,{D^{'}})$ is a Golay complementary pair.

\emph{Type 1}: By $D \star {D^{'}} \leq 2n$, $F \star {F^{'}} \leq 2n$, $G \star {G^{'}} \leq 4n$, \emph{Lemma 2} and  \cite{Li10}, we have
\begin{equation}\label{eq_Ni_ap2_4}
\begin{array}{l}
{\rm{~~}}{\textrm{J}} \star {\textrm{J}^{'}} \\
= \sum\limits_{u = 1-n}^{n - 1} \bigg| {\sum\limits_{i = 0}^{n - u - 1} [a_1^2({\zeta ^{{D_i} - {D_{i + u}}}} + {\zeta ^{{D_i^{'}} - {D_{i + u}^{'}}}})}  \\
+ a_2^2({\zeta ^{{F_i} - {F_{i + u}}}} + {\zeta ^{{F_i^{'}} - {F_{i + u}^{'}}}})  \\
+ a_3^2({\zeta ^{{G_i} - {G_{i + u}}}} + {\zeta ^{{G_i^{'}} - {G_{i + u}^{'}}}})   \\
+ {a_1}{a_2}({\zeta ^{{D_i} - {F_{i + u}}}} + {\zeta ^{{F_i} - {D_{i + u}}}} + {\zeta ^{{D_i^{'}} - {F_{i + u}^{'}}}} + {\zeta ^{{F_i^{'}} - {D_{i + u}^{'}}}})  \\
 + {a_1}{a_3}({\zeta ^{{D_i} - {G_{i + u}}}} + {\zeta ^{{G_i} - {D_{i + u}}}} + {\zeta ^{{D_i^{'}} - {G_{i + u}^{'}}}} + {\zeta ^{{G_i^{'}} - {D_{i + u}^{'}}}})  \\
+ {a_2}{a_3}({\zeta ^{{F_i} - {G_{i + u}}}} + {\zeta ^{{G_i} - {F_{i + u}}}} + {\zeta ^{{F_i^{'}} - {G_{i + u}^{'}}}} + {\zeta ^{{G_i^{'}} - {F_{i + u}^{'}}}})]  \bigg| \\
\leq \sum\limits_{u = 1-n}^{n - 1} \left| \sum\limits_{i = 0}^{n - u - 1} a_1^2({\zeta ^{{D_i} - {D_{i + u}}}} + {\zeta ^{{D_i^{'}} - {D_{i + u}^{'}}}}) \right|
\end{array}
\end{equation}
\begin{equation}\label{eq_Ni_ap2_3}
\begin{array}{l}
+ \sum\limits_{u = 1-n}^{n - 1} \left| \sum\limits_{i = 0}^{n - u - 1} a_2^2({\zeta ^{{F_i} - {F_{i + u}}}} + {\zeta ^{{F_i^{'}} - {F_{i + u}^{'}}}}) \right| \\
+ \sum\limits_{u = 1-n}^{n - 1} \left|  \sum\limits_{i = 0}^{n - u - 1} a_3^2({\zeta ^{{G_i} - {G_{i + u}}}} + {\zeta ^{{G_i^{'}} - {G_{i + u}^{'}}}})\right|   \\
+ {a_1}{a_2} \ast 4n +2\times  {\sum\limits_{u = 1}^{n - 1} \bigg| {\sum\limits_{i = 0}^{n - u - 1} {{a_1}{a_2} \times {\zeta ^{{D_i} - {D_{i + u}}}}} } }   \\
\left.\times [1 + {( - 1)^{{i_{\pi (m - 1)}} - {{(i + u)}_{\pi (m - 1)}}}}] \times ({\zeta ^{ - s_{i + u}^{(1)}}} + {\zeta ^{s_i^{(1)}}})  \right|  \nonumber\\
+ \sum\limits_{u = 1-n}^{n - 1} \left| { {\sum\limits_{i = 0}^{n - u - 1} {{a_1}{a_3} \times {\zeta ^{{D_i} - {D_{i + u}}}}} } } \right. \nonumber\\
\left.\times [1 + {( - 1)^{{i_{\pi (m - 1)}} - {{(i + u)}_{\pi (m - 1)}}}}] \times ({\zeta ^{ - s_{i + u}^{(2)}}} + {\zeta ^{s_i^{(2)}}})  \right| \\
+ \sum\limits_{u = 1-n}^{n - 1}\left| { {\sum\limits_{i = 0}^{n - u - 1} {{a_2}{a_3} \times {\zeta ^{{F_i} - {F_{i + u}}}}} } } \right. \nonumber\\
\left.\times [1 + {( - 1)^{{i_{\pi (m - 1)}} - {{(i + u)}_{\pi (m - 1)}}}}] \times ({\zeta ^{ - s_{i + u}^{(3)}}} + {\zeta ^{s_i^{(3)}}})  \right| \\
\left. + {\zeta ^{{F_i^{'}} - {G_{i + u}^{'}}}} + {\zeta ^{{G_i^{'}} - {F_{i + u}^{'}}}})] \right|   \\
\leq a_1^2 \ast 2n + a_2^2 \ast 2n + a_3^2 \ast 4n + {a_1}{a_2} \ast 4n \approx 3.62n.
\end{array}
\end{equation}

\emph{Type 2}: By $D \star {D^{'}} \leq 2n$, $F \star {F^{'}} \leq 4n$, $G \star {G^{'}} \leq 4n$, \emph{Lemma 3} and  \cite{Li10}, we have
\begin{equation}\label{eq_final}
\begin{array}{l}
{\rm{~~}}{\textrm{J}} \star {\textrm{J}^{'}} \\
= \sum\limits_{u = 1-n}^{n - 1} \bigg| {\sum\limits_{i = 0}^{n - u - 1} [a_1^2({\zeta ^{{D_i} - {D_{i + u}}}} + {\zeta ^{{D_i^{'}} - {D_{i + u}^{'}}}})}  \\
+ a_2^2({\zeta ^{{F_i} - {F_{i + u}}}} + {\zeta ^{{F_i^{'}} - {F_{i + u}^{'}}}})  \\
+ a_3^2({\zeta ^{{G_i} - {G_{i + u}}}} + {\zeta ^{{G_i^{'}} - {G_{i + u}^{'}}}})   \\
+ {a_1}{a_2}({\zeta ^{{D_i} - {F_{i + u}}}} + {\zeta ^{{F_i} - {D_{i + u}}}} + {\zeta ^{{D_i^{'}} - {F_{i + u}^{'}}}} + {\zeta ^{{F_i^{'}} - {D_{i + u}^{'}}}}) \\
 + {a_1}{a_3}({\zeta ^{{D_i} - {G_{i + u}}}} + {\zeta ^{{G_i} - {D_{i + u}}}} + {\zeta ^{{D_i^{'}} - {G_{i + u}^{'}}}} + {\zeta ^{{G_i^{'}} - {D_{i + u}^{'}}}})  \\
+ {a_2}{a_3}({\zeta ^{{F_i} - {G_{i + u}}}} + {\zeta ^{{G_i} - {F_{i + u}}}} + {\zeta ^{{F_i^{'}} - {G_{i + u}^{'}}}} + {\zeta ^{{G_i^{'}} - {F_{i + u}^{'}}}})]  \bigg| \\
 \leq a_1^2 \ast 2n + a_2^2 \ast 4n + a_3^2 \ast 4n  \approx 2.48n.
\end{array}
\end{equation}

In summary, the proposed $64$-QAM sequence $\textrm{H}$ is a near-complementary sequence because $2n \le \textrm{H} \star \textrm{H}^{'} \ll 2{n^2}$. Moreover, according to \eqref{eq pmepr2}, \eqref{eq_Ni_ap2_3} and \eqref{eq_final}, the PMPERs of the $\textrm{H}$ for \emph{Type 1} and \emph{Type 2} satisfy that $\textrm{PMEPR}^1 \leq 3.62 \rm{~} \textrm{and} \rm{~} \textrm{PMEPR}^2 \leq 2.48$, respectively. Then, the proof of \emph{Theorem 2} is complete.

\end{document}